\documentclass{revtex4-1}

\usepackage{graphicx}
\usepackage{color}
\usepackage{amsmath}
\usepackage{float}
\usepackage{tabularx,subfigure}
\usepackage[toc,page]{appendix}
\usepackage{natbib}

\begin{document}

\author{Miguel \surname{Pi\~neirua}}
\email[]{miguel.pineirua@ensta-paristech.fr}
\affiliation{Unit\'e de M\'ecanique (UME), ENSTA-Paristech, 91761 Palaiseau, France}
\author{Olivier \surname{Doar\'e}}
\email[]{olivier.doare@ensta-paristech.fr}
\affiliation{Unit\'e de M\'ecanique (UME), ENSTA-Paristech, 91761 Palaiseau, France}
\author{S\'ebastien Michelin}
\email[]{sebastien.michelin@ladhyx.polytechnique}
\affiliation{LadHyX--D\'epartement de M\'ecanique, \'Ecole Polytechnique, Route de Saclay, 91128 Palaiseau, France}

\title{Influence and optimization of the electrodes position in a piezoelectric energy harvesting flag}

\begin{abstract}
\textit{Fluttering piezoelectric plates may harvest energy from a fluid flow by converting the plate's mechanical deformation into electric energy in an output circuit. This work focuses on the influence of the arrangement of the piezoelectric electrodes along the plate's surface on the energy harvesting efficiency of the system, using a combination of experiments and numerical simulations. A weakly non-linear model of a plate in axial flow, equipped with a discrete number of piezoelectric patches is derived and confronted to experimental results. Numerical simulations are then used to optimize the position and dimensions of the piezoelectric electrodes. These optimal configurations can be understood physically in the limit of small and large electromechanical coupling.}
\end{abstract}


\maketitle

%%%%%%%%%%%%%%%%%%%%%%%%%%%%%%%%%%%%%%%%%%%%%%%%%%%%%
\section{Introduction}

In recent years, a significant research effort has been dedicated to the development of new flow energy-harvesting systems and the characterization of their efficiency. Many of such systems are based on flow-induced vibrations of structures, such as vortex-induced vibrations \cite{taylor2001,bernitsas2008,grouthier2013omae}, flutter of wing profiles \cite{erturk2010,bryant2011,xiao2014}, of cylinders in axial flows \cite{singh2012,singh2012b}, or flags \cite{tang2009b,doare2011b,michelin2013,akcabay2012}. The present work focuses on the latter, and considers a flag covered with piezoelectric patches that convert mechanical deformation into electric charge transfer inside an output circuit.

In this article the term {\em flag} refers in fact to a clamped-free plate in an axial flow. It is well-established that such structure can undergo strong self-sustained oscillations once a critical flow velocity is overpassed (see the recent review by Ref.~\cite{shelley2011} and references therein). These oscillations originate from an instability of the equilibrium position, resulting from the interaction of the flow forces with the solid's inertia and rigidity. This instability can be predicted from the linearized dynamics equations and different methods and models have been proposed depending on the aspect ratio of the flag, ranging from two-dimensional models (large span, \cite{kornecki1976,guo2000}) to slender body models (short span, \cite{lemaitre2005}). The general case of a three dimensional fluid flow around a rectangular plate showed the limits of applicability of both models \cite{eloy2007,doare2011a}. Linear analysis is however not able to predict the amplitude of flapping in the saturated regime, hence it is necessary to consider the nonlinear dynamics of the flag to be able to determine how much energy can be harvested. Beyond full numerical simulations of the coupled fluid-solid problems, simplified fluid-solid nonlinear models have been proposed in the limit of two-dimensional flows \cite{alben2009,michelin2008} or slender body problems \cite{eloy2012,singh2012b}. 

Piezoelectric patches  have traditionally been studied to couple structural mechanics to electrical circuits in the context of passive damping of structural vibrations \cite{hagood1991}, of active control of vibrations \cite{preumont2002} and of course energy harvesting \cite{lefeuvre2006,anton2007,elvin2013,chen2013}. The use of a piezoelectric flag to convert the kinetic energy of a flow was initially introduced by Ref.~\cite{taylor2001}, to produce energy from the flapping of a flexible membrane forced by the wake of an upstream obstacle. In the last five years, several experimental, numerical and theoretical studies have been conducted to study the possibility to exploit the flutter instability itself using piezoelectric plates \citep{doare2011b,dunnmon2011,akcabay2012b,michelin2013}. In these works, the plate was either entirely or partially covered by one piezoelectric pair \cite{akcabay2012b,dunnmon2011}, or completely covered by a large amount of small piezoelectric elements, so that a continuous model for a homogenous plate can be considered \cite{doare2011b,michelin2013}. In the latter case and for purely resistive output circuits, it was shown that the maximum efficiency is obtained for large mass ratio when the timescale of the electrical circuit is equal to the fluid-solid instability timescale \cite{michelin2013}. For practical applications, however, only a finite number of piezoelectric elements can realistically be used. The goal of the present article is therefore to provide some understanding on the impact of such a discrete coverage on the dynamics and performance of the system, a critical question that has remained so far unanswered.

In contrast with simplified representations of the energy harvesting process (e.g. pure damping, \cite{tang2009}), such an approach provides a fully-coupled description of the fluid-solid-electric problem, taking into account both the transfer of energy from the solid motion to the output circuit and the feedback effect of the piezoelectric patches on the flag's dynamics. The nonlinear dynamics of the flag (even in the absence of any piezoelectric coupling) exhibit strong inhomogeneities  of the local deformation and motion along the plate: near the clamped leading edge, the flag has little displacement but significant curvature, while near the free trailing edge, the flag has zero curvature but undergoes large displacements. Local stretching and compression of the piezoelectric patches resulting from the flag deformation are responsible for the charge transfer in the output circuit; the location along the structure of the device converting mechanical to electrical energy is therefore expected to strongly affect the efficiency of the system. A similar question was recently addressed in Refs.~\cite{singh2012,singh2012b} in the case of a pure damping model. 

The present work focuses on the optimization of the distribution of a small set of piezoelectric elements (one, two or three pairs) positioned on a plate in an axial flow. In section \ref{sec:model}, the physical model and equations are presented for the fluid-solid-electric system with a finite number of discrete piezoelectric patches.  A method of simulation based on the weakly nonlinear form of these equations is presented that is appropriate to obtain the nonlinear dynamics near the instability threshold. The influence of the number of piezoelectric elements on the stability of the system is investigated in section \ref{sec:stability}. In section \ref{sec:geom}, the impact of this arrangement on the nonlinear dynamics and energy harvesting efficiency is studied using experiments and numerical simulations. Finally, a parametric numerical study is performed to determine the optimal arrangements in the case of one, two or three piezoelectric elements.

%%%%%%%%%%%%%%%%%%%%%%%%%%%%%%%%%%%%%%%%%%%%%%%%%%%%%
\section{\label{sec:model}Physical Model}

\noindent The system considered in this work consists of an elastic flexible plate of length $L$ and width $H$ immersed in an axial flow of uniform density $\rho$ and velocity $U_ \infty$. The plate is inextensible and clamped at its leading edge. For simplicity only purely two-dimensional motions of the plate are considered. The plate's geometry is therefore completely determined by the local orientation of the plate with respect to the flow, $\theta(S,T)$, where $S$ is the curvilinear coordinate along the plate and $T$ is time. 
\begin{figure}[h]
\centering
\includegraphics[width=\textwidth]{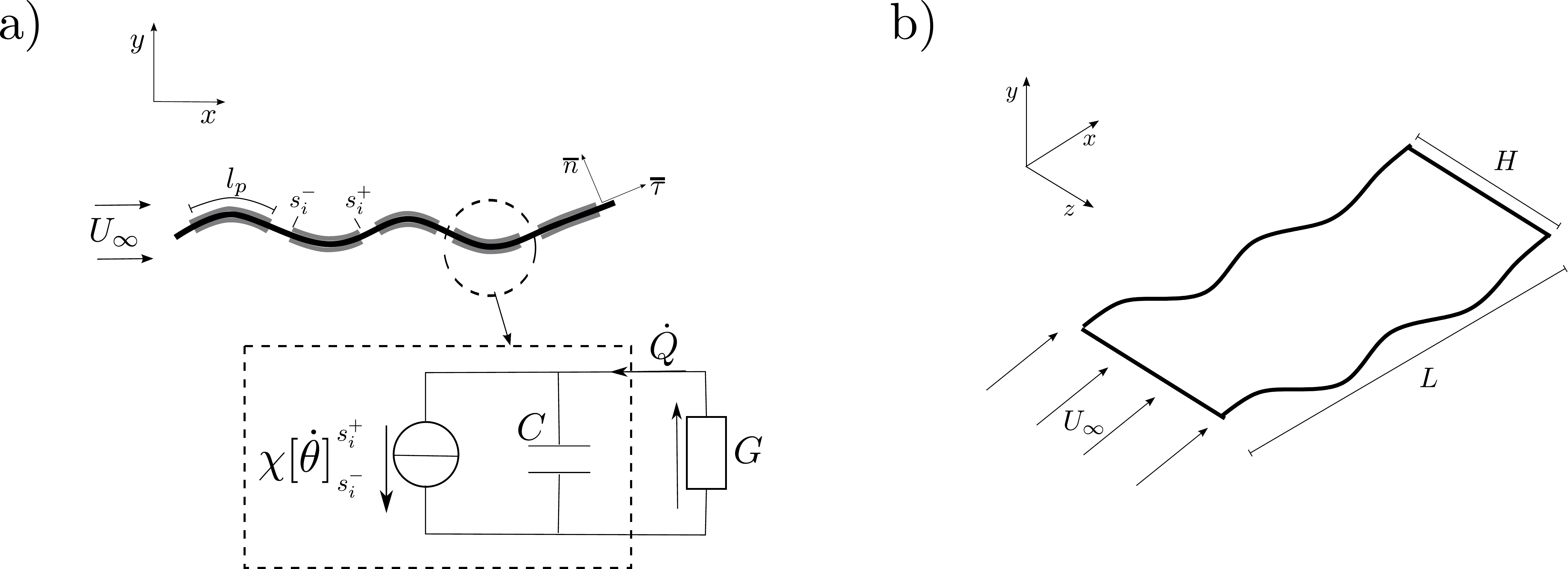} 
 \caption{(a) Two dimensional flapping of a flexible plate covered by pairs of piezoelectric patches. (b) Flexible plate flapping in uniform axial flow.}
\label{esquema}
 \end{figure}
A finite number $N_p$ of piezoelectric electrodes of length $L_{pi}$ and width $H$ are attached on each side of the plate. For each piezoelectric patch pair, the negative electrodes are shunted across the plate while their positive electrodes are connected to an external circuit. The charge (per unit length in the streamwise direction) in each piezo pair is given by \cite{doare2011b}:
\begin{equation}
Q_i=\frac{\chi}{L_{pi}}[\theta]_{S_i^-}^{S_i^+}+CV_i\;,
\label{eq_piezo}
\end{equation}
with $V_i$ the voltage between the positive electrodes of the \textit{i}th piezo pair whose left and right edges are positioned at ${S_i^-}$ and ${S_i^+}$ respectively. $C=C_i/L_{pi}$ is the equivalent capacitance per unit length of the piezo pair and $\chi$ is the mechanical/piezoelectrical conversion factor of the piezoelectric material. Considering that each electrode is connected to a purely resistive circuit, Ohm's law leads to:
\begin{equation}
 \frac{\partial Q_i}{\partial T}+GV_i=0,
 \label{eq_circuit}
\end{equation}
where $G=G_i/L_{pi}=1/R_iL_{pi}$ is the conductivity per unit length of the harvesting circuit, with $R_i$ the circuit's resistance.

The voltage $V_i$ between the positive electrodes also generates an internal torque in the piezoelectric patches, and thus on the flexible plate, so that the total internal torque in the plate  and piezoelectric patches assembly is obtained as
\begin{equation}\label{eq:moment}
M=B\frac{\partial \theta}{\partial S}-\chi \sum_{i=1}^{N_p} V_i F_i,
\end{equation}
 where $B$ is the flexural rigidity, which depends on the respective thickness, Young's modulus and Poisson's ratio of the two constitutive materials \citep{doare2011b}. In equation \eqref{eq:moment}, $F_i(S)$ is the polarization function of the \textit{i}th patch. In the present approach, $F_i(S)=H_s(S-S_i^-)-H_s(S-S_i^+)$, with $H_s$ the Heaviside step function. The conservation of momentum then leads to \cite{michelin2013}:
\begin{equation}
 \mu\frac{\partial^2 \mathbf{X}}{\partial T^2}=\frac{\partial}{\partial S}\left(F_t\boldsymbol{\tau}-\frac{\partial}{\partial S}\left(B\frac{\partial \theta}{\partial S}-\chi \sum_{i=1}^{N_p} V_i F_i\right)\mathbf{n}\right)-P\mathbf{n} ,
 \label{eq_plaque}
\end{equation}
where $\mathbf X(S,T)$ is the local plate position and $\mu$ is the  mass per unit length of the assembly.  $F_t$ is the internal tension in the plate that enforces inextensibility and $P$ is the pressure force density exerted by the surrounding fluid over the plate. In Eq.~\eqref{eq_plaque}, $\mathbf{n}$ and $\mathbf{\tau}$ are the unit normal and tangent vectors to the plate (see Figure \ref{esquema}). Following previous works on  flapping flag modeling \cite{eloy2012,singh2012b}, $P$ is decomposed into a reactive part $P_{\text{reac}}$ accounting for potential flow effects (e.g. added mass), corrected by a resistive part $P_{\text{res}}$ accounting for the form drag on the plate resulting from lateral flow separation:
\begin{equation}
 P=P_{\text{reac}}+P_{\text{res}}=m_a\rho H^2\left(\dot{w}-(wu)'+\frac{1}{2}w^2 \kappa\right)+\frac{1}{2}\rho C_d H |w| w,
\end{equation}
where $m_a$ is the added mass coefficient of the transverse section ($m_a=\pi/4$ for a rectangular plate \cite{blevins1990}), and $C_d=1.8$ the drag coefficient \cite{buchak2010}.  $w$ and $u$ are respectively the normal and longitudinal components of the plate's velocity relative to the fluid flow, such that $\dot{\mathbf{X}}-U_\infty \mathbf{e}_x=u\boldsymbol{\tau}+w\mathbf{n}$, with $\mathbf{e}_x$ the unitary vector in the $x$ direction.

In the present approach, it is assumed that the electrode does not affect the density and bending rigidity of the sandwich plate that are considered homogeneous along the assembly (see \citep{thomas2009} for a study on the passive damping of vibrating beams including this effect). Also, in the approach followed in this work, $P_{\text{res}}$ accounts for the effect of the lateral flow detachment in the form of a drag. Flow detachment from the trailing edge and wake dynamics are only accounted through the reactive term which models the advection of added fluid momentum (see Ref.~\cite{lighthill1971} for more details). Other models exist that consider the influence of an unsteady wake \cite{howell2009,alben2009}.

% ------------------------------------
\subsection{Non-dimensional equations}

\noindent Using $L$, $L/U_\infty$, $U_\infty\sqrt{\mu/c}$ and  $U_\infty\sqrt{\mu c}$ as characteristic length, time, voltage and charge respectively, equations \eqref{eq_piezo}--\eqref{eq_plaque} become in non-dimensional form:
\begin{align}
q_i & =v_i+\frac{\alpha}{U^* \gamma_i}[\theta]_{s_i^-}^{s_i^+},\label{eq_piezo_adim}\\
\beta\frac{\partial q_i}{\partial t}+v_i &= 0,\label{eq_circ_adim}\\
\frac{\partial^2 \mathbf{x}}{\partial t^2} & = \frac{\partial}{\partial s}(f_\tau\boldsymbol{\tau})-\frac{1}{U^{*2}}\frac{\partial^ 2}{\partial s^2}\left(\frac{\partial\theta}{\partial s}\mathbf{n}\right)-\left(M^* p_{\text{res}} +M^* H^*m_ap_{\text{reac}}\right)\mathbf{n} \nonumber \\
& \;\;\;\; +\frac{\alpha}{U^*}\sum_i v_i\left[\delta'(s-s_i^-)-\delta'(s-s_i^+)\right]\mathbf{n},\label{eq_plaque_adim}
\end{align}
with clamped-free boundary conditions. The problem is characterized by the following non-dimensional parameters:
\begin{align}
&\text {- the non dimensional velocity}\quad U^*=LU_\infty\sqrt{\frac{\mu}{B}},\\ 
&\text {- the electro-mechanical coupling factor}\quad\alpha=\frac{\chi}{\sqrt{BC}},\\ 
&\text {- the fluid-solid inertial ratio}\quad M^*=\frac{\rho L H}{\mu},\\ 
&\text {- the non dimensional plate span}\quad H^*=\frac{H}{L},\\ 
&\text {- the non dimensional electrode length}\quad\gamma_i=\frac{L_{pi}}{L},\\ 
&\text {- the tuning coefficient of the fluid-solid and electric systems}\quad \beta=\frac{U_\infty C}{L G}.
\end{align}

% --------------------------------------------------
\subsection{Weakly non-linear form of the equations}

\noindent Equation (\ref{eq_plaque_adim}) is projected onto the $x$ and $y$ directions in order to obtain two equations for $x(s,t)$ and $y(s,t)$ respectively.  The horizontal projection is used to eliminate the tension term $f_\tau$ from the $y$ projection.  Finally, $x$ and its derivatives are eliminated using the inextensibility condition.  Keeping terms up to $O(y^3)$ one obtains a weakly non-linear equation for $y(s,t)$ (see also Ref.~\citep{eloy2012} for the case of a non-piezoelectric plate):
\begin{equation}
L(y)+f_\text{m}(y)+\frac{1}{U^{*2}}f_\text{B}(y)-\frac{\alpha}{U^*}f_\chi(y,v)+M^* f_{\text{res}}(y)+M^* H^*m_a f_{\text{reac}}(y)=0,
\label{ydest}
\end{equation}
where 
\begin{align}
&L(y)=\ddot{y}+\frac{1}{U^{*2}}y^{(4)}-\frac{\alpha}{U^*}\sum_i v_i(\delta'(s-s_i^-)-\delta'(s-s_i^+))+M^*H^*(y''+2\dot{y'}+\ddot{y}),\label{eq:linear}\\
&f_\text{m}(y)=y'\int_0^s(\dot{y'^2}+y'\ddot{y'})ds-y''\int_s'^1\int_0^s(\dot{y'^2}+y'\ddot{y'})dsds',\label{eq:nl1}\\
&f_\text{B}(y)=4y'y''y'''+y'^2y^{(4)}+y''^3,\\
&f_\chi(y,v_1..{v_{N}}_p)=\frac{1}{2}y'^2\sum_i v_i(\delta'(s-s_i^-)-\delta'(s-s_i^+))+y'y''\sum_i v_i(\delta(s-s_i^-)-\delta(s-s_i^+)),\\
&f_{\text{res}}(y)=\frac{1}{2}C_D|y'+\dot{y}|(y'+\dot{y}),\\
&f_{\text{reac}}(y)=-\frac{1}{2}y''y'^2+\dot{y'}y'^2-3y''y'\dot{y}-2\dot{y'}y'\dot{y}-\frac{1}{2}y''\dot{y}^2+y'\int_0^s(\dot{y'}^2+y'\ddot{y'})ds\nonumber \\&+2(y''+\dot{y'})\int_0^s\dot{y'}y'ds-y''\int_s^1y'(y''+2\dot{y'}+\ddot{y})ds.\label{eq:nl5}
\end{align}
Equation~\eqref{eq:linear} corresponds to the linearized dynamics while Equations~\eqref{eq:nl1}--\eqref{eq:nl5} correspond to nonlinearities related to inertia, stiffness, piezoelectric coupling, resistive and reactive flow effects, respectively. 

% ---------------------------
\subsection{Numerical method}

\noindent A Galerkin decomposition is used to solve Equation~\eqref{ydest}: the vertical displacement $y$ is expanded as a superposition of clamped-free beam eigenmodes $\phi_p(s)$,
\begin{equation}
 y(s,t)=\sum_{p=1}^\infty X_p(t)\phi_p(s).
\end{equation}
Next, equation (\ref{ydest}) is projected on the same set of eigenmodes. Details of the Galerkin projection can be found in \ref{Galerkin-projection}. After truncation to $N$ linear modes, the resulting coupled system of equations is integrated numerically using a semi-implicit step-adaptive fourth order Runge-Kutta method.

% -----------------------------------------------------
\subsection{Harvested energy and harvesting efficiency}

\noindent Once the dynamics of the piezoelectric flag is obtained using the method above and a limit-cycle oscillation is reached, the total energy harvested by the system is computed as the energy dissipated in all the resistive circuits. In dimensional form, this reads as
 \begin{align}
 P_e=\left<\sum_{i=1}^{N_p}G_iV_i^2\right>,
 \end{align}
 with $\left<.\right>$ the temporal average over one oscillation period. The harvesting efficiency $\eta=P_e/P_f$ is then computed as the fraction of the fluid kinetic energy flux $P_f$ through the vertical cross-section occupied by the flag that is actually harvested and dissipated in the output circuits. Using the non-dimensional variables defined above, this can be written as
 \begin{equation}
\eta=\frac{1}{M^* \beta A^*}\left<\sum_{i=1}^{N_p}\gamma_i v_i^2\right>,
\label{efficacite}
\end{equation}
with $A^*=A/L$ and $A$ the flapping amplitude of the flag.

%%%%%%%%%%%%%%%%%%%%%%%%%%%%%%%%%%%%%%%%%%%%%%%%%%%%%
\section{Flapping Instability : linear analysis}

\label{sec:stability}
\begin{figure}
\centering
 \begin{tabular}{l@{}l}
 (a)&(b)\\
\includegraphics[width=0.5\textwidth]{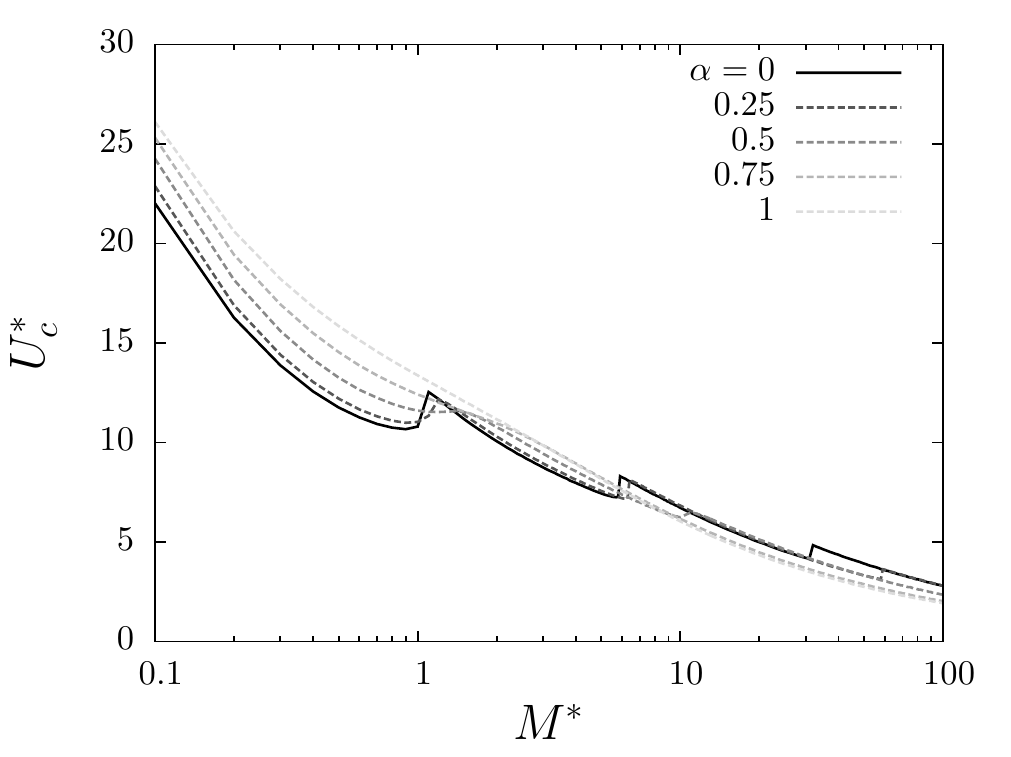}&
\includegraphics[width=0.5\textwidth]{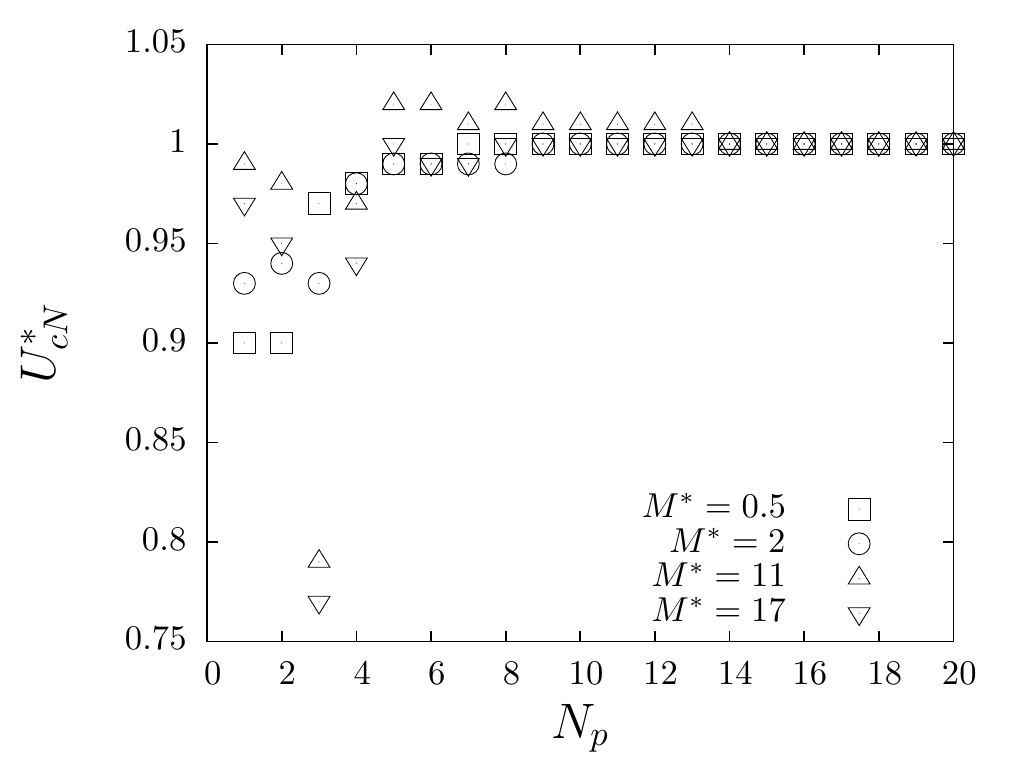}
 \end{tabular}
 \caption{(a) :  Evolution of the stability threshold as a function of $M^*$ for different values of the coupling factor $\alpha$, $\beta=1=cst$, $N_p=100$.  (b) : Impact of the number of electrodes on the critical stability threshold.  Values of $U^*_c $ are normalized using the critical value corresponding to $N_p=100$.}
  \label{lin-stab-an}
\end{figure}

\noindent As a first validation of our model, we investigate the effect of the piezoelectric arrangement of the flag on its stability for a continuous coverage ($s_{i+1}^-=s_i^+$), and a large number of piezoelectric patches. Beyond a critical value $U^*_c$ of the flow velocity the system  becomes unstable to flutter \citep{shelley2011}. Figure \ref{lin-stab-an} (left) shows the critical velocity $U_c^*$ as a function of the mass ratio $M^*$ for different values of the coupling coefficient and a large number of small piezoelectric patches ($N_p=100$).  Without piezoelectric coupling, $U_c^*$ is a decreasing function of $M^*$. Cusps in the curve represent a switch in the mode becoming unstable at the lowest flow velocity (see for example \cite {eloy2007}). These results match that obtained in Ref.~\cite{michelin2013} using a continuous model, thus indicating that for a large amount of piezoelectric patches, the present discrete model indeed converges to the continuous framework.

The electro-mechanical coupling also clearly influences the stability threshold of the flag. In the coupled case ($\alpha\neq0$) and for small values of $M^*$, the damping induced by the electro-mechanical coupling increases the critical velocity $U^*_c$. This effect is more important as the coupling factor $\alpha$ is increased. However, for high values of $M^*$, the electro-mechanical coupling has a destabilizing  effect, which was interpreted in Ref.~\cite{doare2011b} in terms of negative energy waves present on the non-dissipative flag \cite{benjamin1963,cairns1979,doare2010}.

Several previous studies have focused on the limit of the continuous coverage of the plate by infinitesimal patches ($N_p\gg 1$, \citep{bisegna2006,doare2011b,michelin2013}). The discrete approach considered here allows us to investigate the convergence of the results with $N_p$ and the applicability of the continuous approach to experimental situations where a finite and often small number of patch pairs is considered. Figure~\ref{lin-stab-an} (right) indeed shows a rapid convergence of the critical velocity for $N_p>12$, beyond which the results of the continuous model are recovered \citep{bisegna2006,doare2011b,michelin2013}. Moreover, it also shows that using fewer patches can significantly modify the stability threshold, positively or negatively, which motivates the present study on a finite number of piezoelectric pairs. 

The rapid convergence of the results beyond $N_p\approx 10$ can be understood as follows: for the values of the mass ratio considered, the dominant unstable mode is of low order resulting in typical wavelengths greater than half a flag length. For a large enough number of patches, the patch length is small enough that the entire mode structure is well captured and the discrete nature of the piezoelectric coverage has no influence.

%%%%%%%%%%%%%%%%%%%%%%%%%%%%%%%%%%%%%%%%%%%%%%%%%%%%%
\section{\label{sec:geom}Optimization of the electrodes' position}

During its self-sustained flapping motion, the flag undergoes complex and non-uniform deformations. Electric charge transfer between the patches' electrodes is intimately linked to the local deformation of the flag. An important challenge for maximizing the amount of energy harvested by such a device therefore lies in the correct positioning of the electrodes. When the electromechanical coupling is large, the piezoelectric patches can significantly modify the flag's dynamics resulting in a non-trivial optimization process. 

In order to address this question, we first present  a set of preliminary experiments exploring the impact of electrode size and position on the global energy harvesting efficiency of the system.  Experimental results are then compared to numerical simulations based on the physical model presented in section \ref{sec:model}.  In the last part of this section, a parametric study is performed based on numerical simulations using more complex geometrical configurations, to identify optimal positioning of the electrodes.  

% -----------------------------------------------------------------------------------------
\subsection{\label{sec:expe} Experimental study of the role of electrode size and position}

\begin{figure}[h]
\centering
 \begin{tabular}{l@{\hspace{0.1\textwidth}}l}
 (a)&(b)\\
\includegraphics[width=0.45\textwidth]{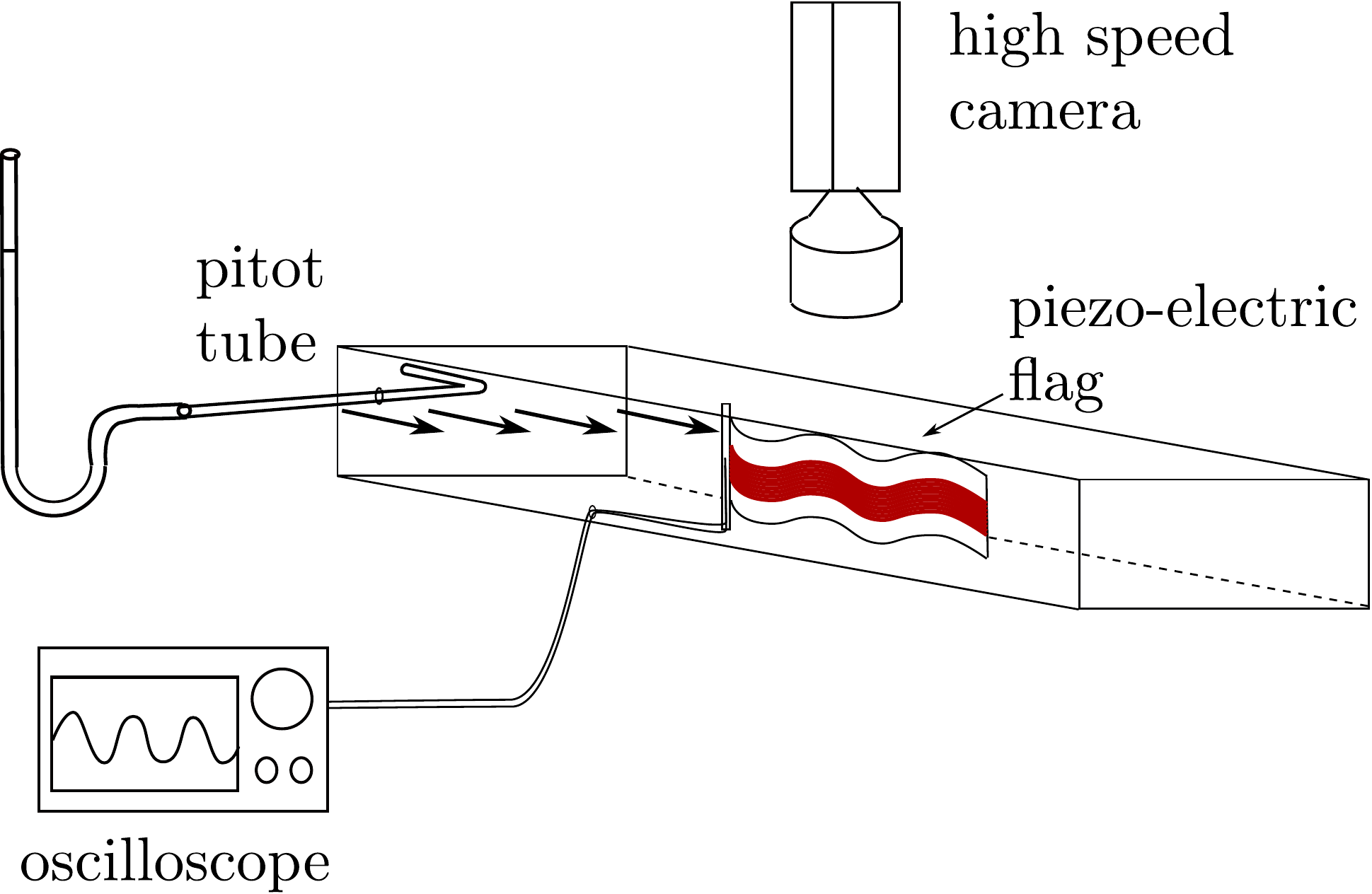}&
\includegraphics[width=0.25\textwidth]{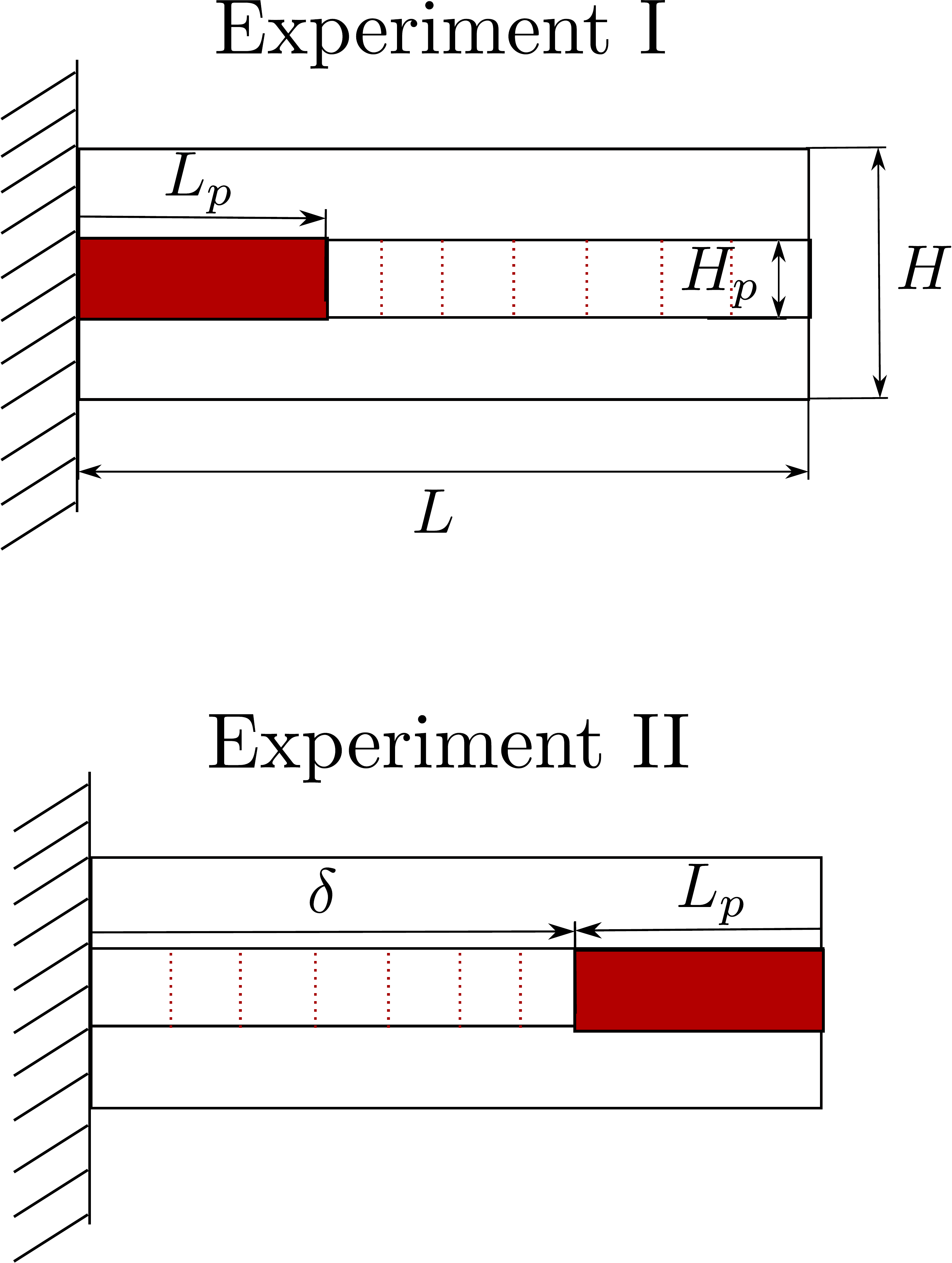}\\
\multicolumn{2}{c}{\vspace{0.5cm}}\\
(c)&(d)\\
\includegraphics[width=0.45\textwidth]{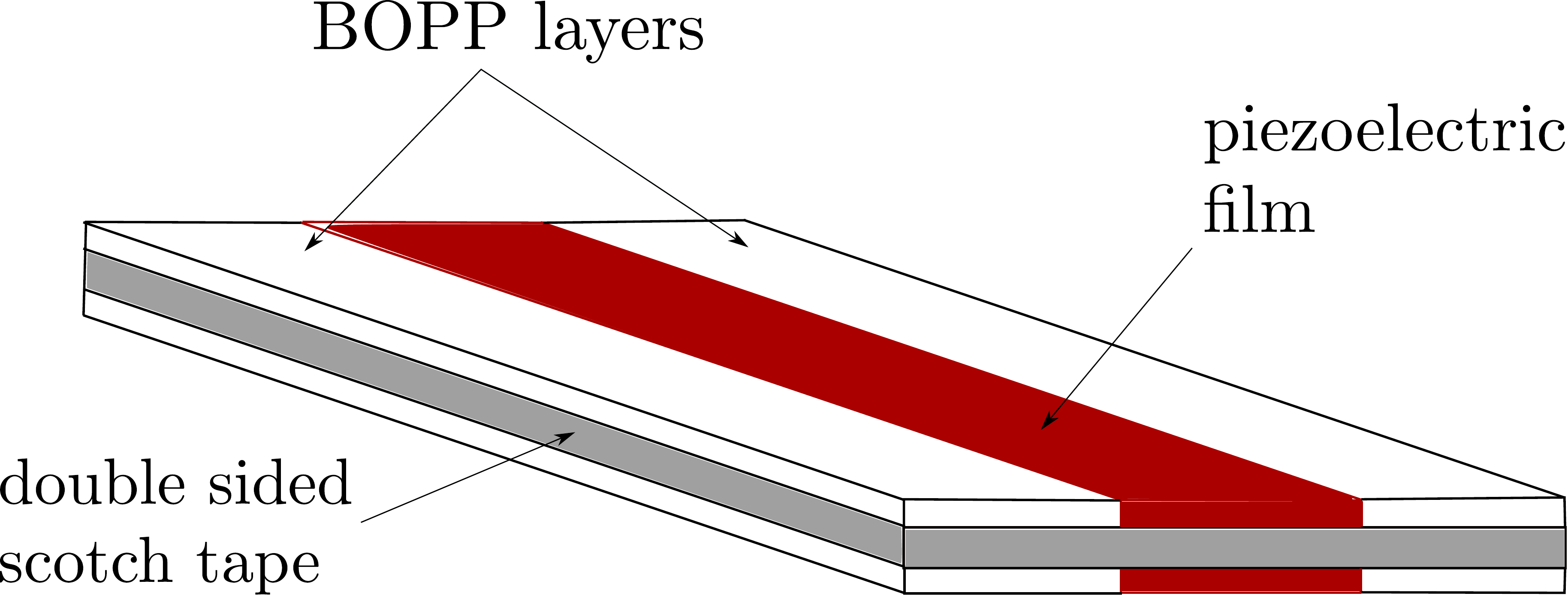}&
\includegraphics[width=0.45\textwidth]{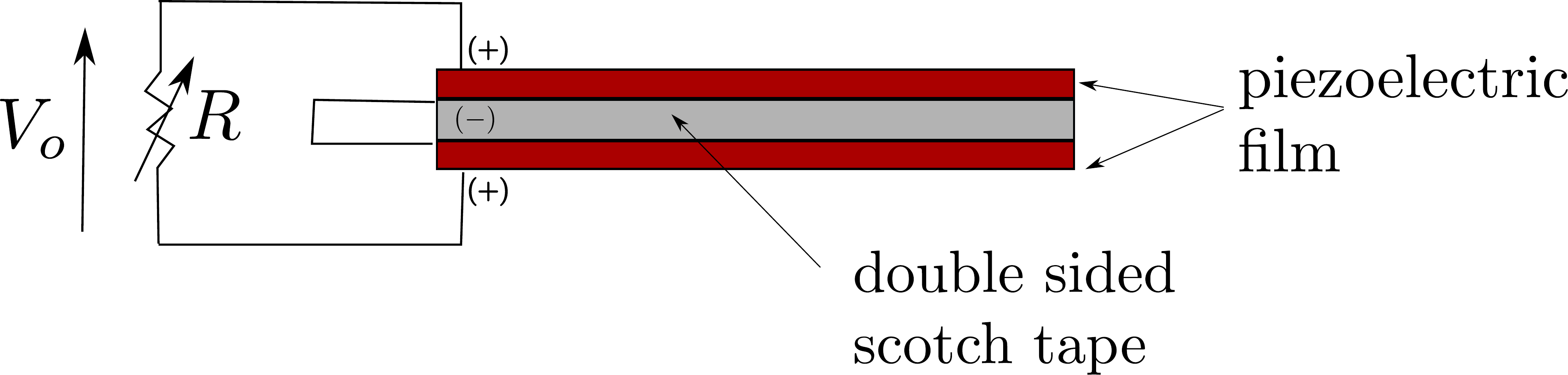}\\
 \end{tabular}
 \caption{(a) Description of the experimental setup. (b) Graphic representation of the electrode size and distribution for experiments 1 and 2.  (c) Schematic of the multilayer composition of the piezoelectric flag.  (d) Connection diagram of the harvesting circuit.}
  \label{dispo_exp}
\end{figure}

The aim of the experiments described below is to obtain some preliminary evidence of the impact of size and position of the electrodes on the harvested energy.  For this purpose we have focused on a simple configuration consisting of a flexible piezo-electric plate covered by a single pair of electrodes of variable length and position, immersed in an axial wind flow of speed $U_\infty$ (figure \ref{dispo_exp}-a).  Two different experimental configurations are considered (see figure \ref{dispo_exp}-b for a graphical description of the experiments): 
\begin{itemize}\renewcommand{\labelitemi}{$-$}
\item{\emph{Experiment 1}:  the length of the electrode $L_p$ is varied while one of its extremities is fixed at the clamped edge of the plate,}
\item{\emph{Experiment 2}:  the length of the electrode is also varied but with one if its extremities fixed to the free edge of the plate.}
\end{itemize}

In these experiments, the  plates include a center core consisting of a double sided adhesive tape of length  $L=12$ cm, width $H=2.5$ cm and thickness $h_s=100$ $\mu$m. Both sides of the tape are  partially covered by a piezoelectric PVDF film of the length $L$  and width $H_p=0.3H$, with its negative polarity facing inwards.  The thickness of the piezoelectric films is $h_p=40$ $\mu$m for experiment 1 and $h_p=50$ $\mu$m for experiment 2. The remaining free surface at both sides of the scotch tape is covered by a layer of \textit{Innova} biaxially oriented polypropylene (BOPP) film of exactly the same thickness as that of the piezoelectric film. Figure \ref{dispo_exp}-c shows a schematic view of the plate composition.  The piezoelectric PVDF films are covered with an external Cr/Au which serves as the electrode. In both experiments this Cr/Au layer is sectioned in order to adjust the desired value of $\delta$ and $L_p$.  The objective of covering only one third of the width of the plate with piezoelectric film is to reduce the impact of the sectioning process on the global rigidity of the plate.  
\begin{figure}[t]
\centering
 \begin{tabular}{l l}
 (a)&(b)\\
 \includegraphics[height=0.3\textwidth]{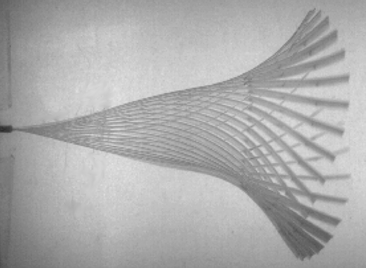}&
 \includegraphics[height=0.3\textwidth]{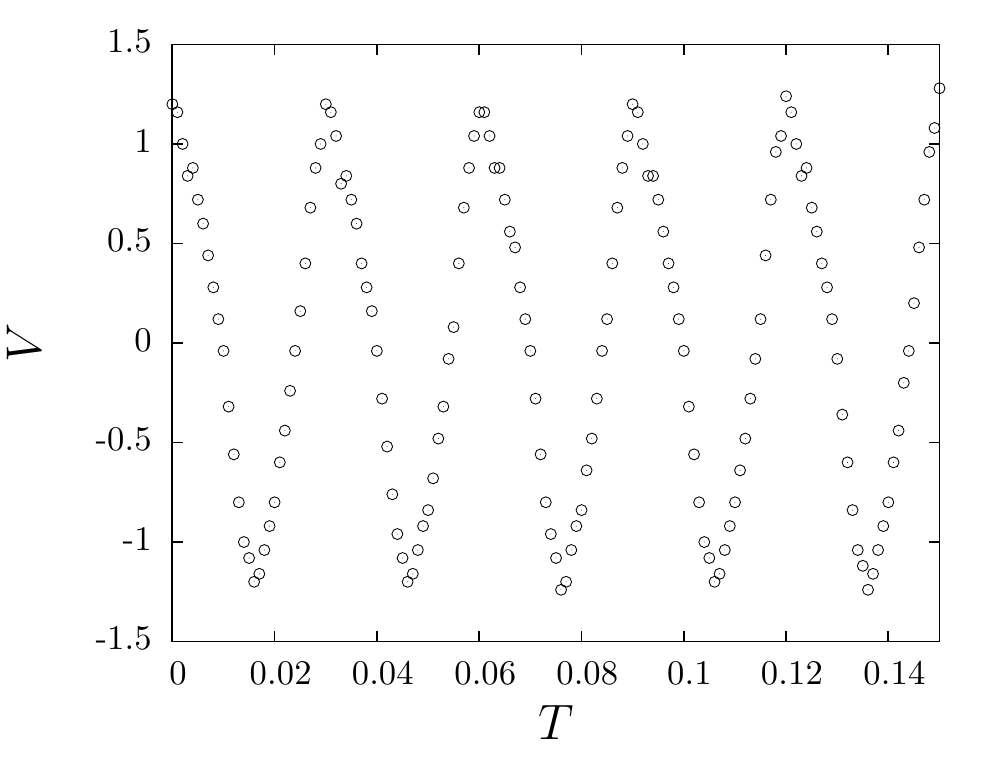}\\
 \hspace{2.3cm}(c)\\
 \multicolumn{2}{c}{ \includegraphics[width=0.55\textwidth]{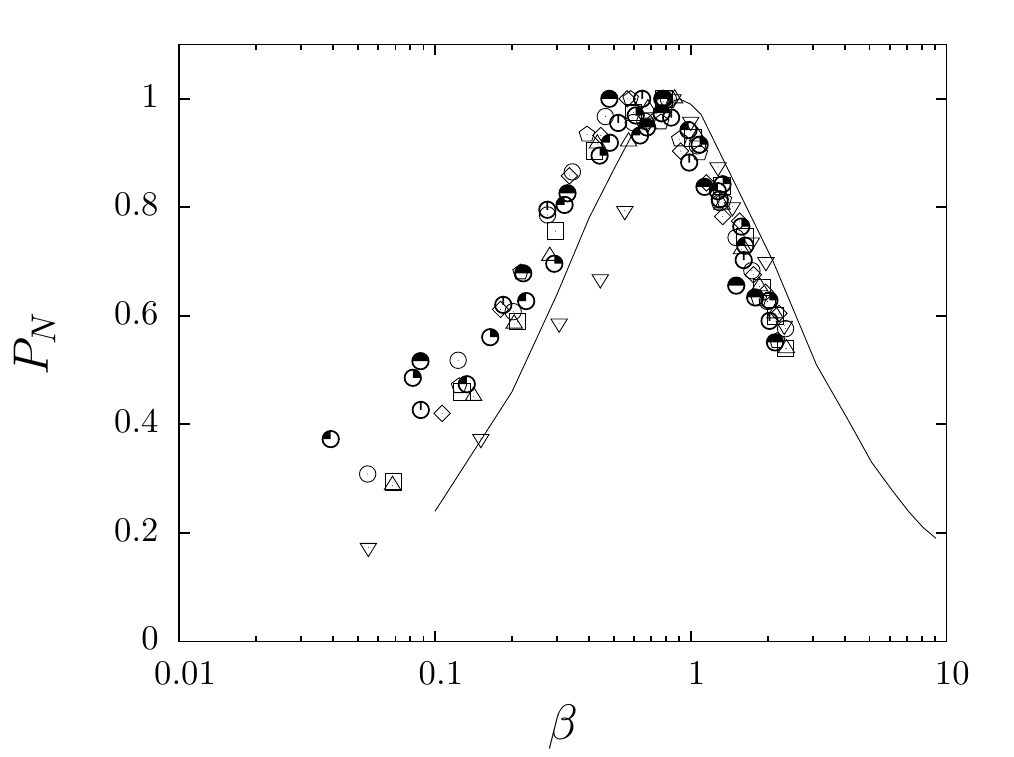}}\\
\end{tabular}
\caption{Top left: Image sequence of the flag deformation over one oscillation period.  Top right : Voltage signal generated by the piezoelectric patches during a typical experiment.  Bottom : Normalized dissipated power $P_N$ as a function of the tuning parameter $\beta$.  Data corresponds to experiment 1, different symbols correspond to different values of $L_p$. The normalized power $P_N$ is calculated as $P_N=P/P_{max}$ for each given value of $L_p$.  The solid line corresponds to a numerical simulation using $L_p=1$. }
\label{VetPot}
\end{figure}

For each experiment and a given electrode length and position, the experimental procedure is as follows : the piezoelectric plate is placed in a wind tunnel of rectangular transversal section 10 cm wide and 5 cm high with transparent walls that allow visual access from the outside \cite{doare2011d}. The negative electrodes of the piezoelectric films are shunted while a resistance $R$ is connected between the positive electrodes (see figure \ref{dispo_exp}-d).  Once the system is set we proceed to generate a wind flow inside the tunnel.  Initially at rest, the wind velocity is gradually increased until  the plate starts to flap in a self sustained oscillatory regime (upper left image in figure \ref{VetPot}), at which point the voltage $V_o$ generated along  the resistance $R$ is measured (figure  \ref{VetPot}, upper right). The power dissipated in the resistance is then calculated as the temporal average  $P_e=<V_o^2/R>$, for a 5 second time frame. The resistance $R$ is varied in order to find the value of $\beta$ that maximizes the dissipated power. Indeed, it is expected that the harvesting efficiency or damping presents a maximum between the extreme values $\beta=0$ (short circuit condition) and $\beta=\infty$ (open circuit condition). The value of $\beta$ maximizing the efficiency corresponds to a tuning of the flapping frequency to the characteristic timescale of the RC circuit consisting of the piezoelectric capacitance and output resistor \citep{michelin2013}. Figure \ref{VetPot} (bottom) shows a good agreement between experiments and simulations obtained from the model described in section \ref{sec:model}. The discrepancies observed  between the experimental and numerical results are mainly  due to small differences in the flapping frequency of the flag, mainly caused by the error in the estimated value of the plate rigidity $B$ used in the simulations. It is important to remark that the optimal $\beta$ value remains fairly constant regardless of the size and position of the electrodes, a direct consequence of the flapping frequency remaining unchanged when changing the electrode position and size. This result was also confirmed in our numerical simulations. 
 
Once the optimal value of $\beta$ is identified, synchronized voltage measurements and image capturing of the flapping plate are carried out using an oscilloscope and a high speed camera placed over the wind tunnel.  Finally, the efficiency of the system is computed as $P_e/P_f$ (see section~\ref{sec:model}).
 
\begin{figure}[t]
\centering
 \begin{tabular}{l l}
\subfigure[Experiment 1]{\includegraphics[width=0.47\textwidth]{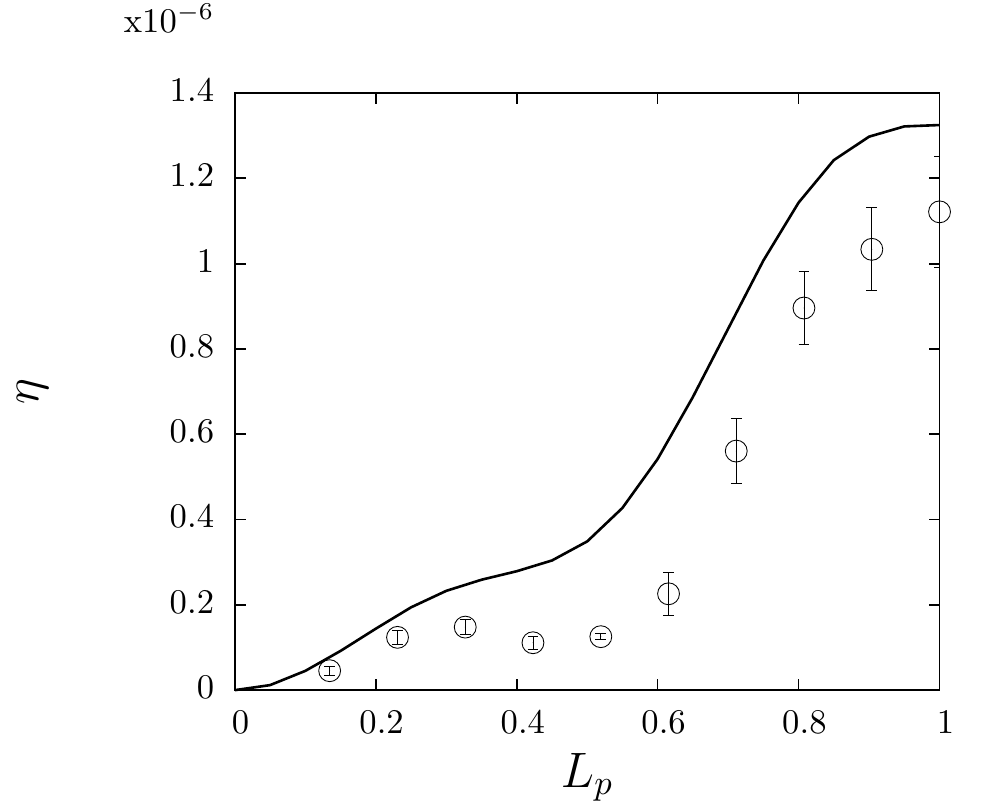}}&
\subfigure[Experiment 2]{\includegraphics[width=0.47\textwidth]{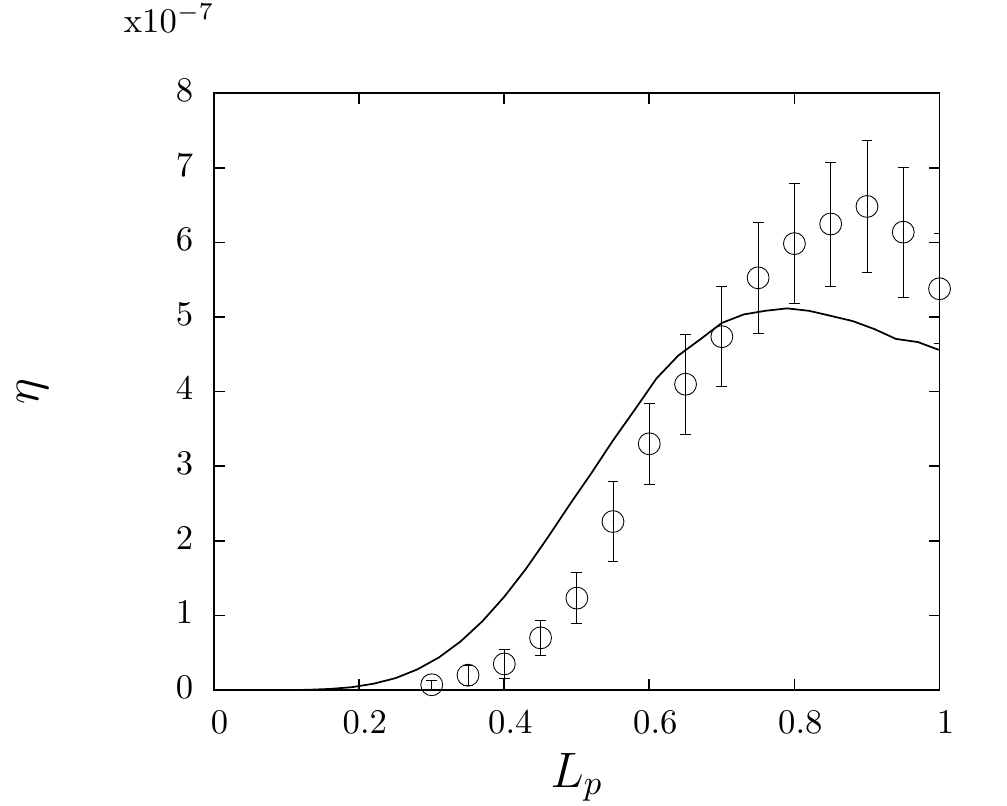}}
 \end{tabular}
 \caption{(a) : Harvesting efficiency as a function of the electrode length $L_p$ for experiment 1.  Experimental parameters are : $U^*=14.6$, $M^*=0.68$, $H^*=0.3$, $\beta=0.8$ and $\alpha=0.073$.  (b) : Harvesting efficiency as a function of the electrode length $L_p$ for experiment 2.  Experimental parameters are : $U^*=15.7$, $M^*=0.61$, $H^*=0.3$, $\beta=0.8$ and $\alpha=0.062$. Error bars correspond to the error propagation of the standard deviation of $V_o$ used to calculate the dissipated power $P_e=<V_o^2/R>$ during a 5 second time frame.  Solid line corresponds to numerical simulations.}
  \label{res_exp}
\end{figure}

Figure \ref{res_exp} shows the evolution of the harvesting efficiency of the piezoelectric flag with the electrode's length $L_p$ for both experiments 1 and 2. A good qualitative agreement is observed between the experimental results and the numerical simulations corresponding to the same non-dimensional parameters. For experiment 1 (left panel), the maximum efficiency is achieved when the electrode occupies the whole length of the plate. For experiment 2 (right panel), the maximum efficiency is achieved when the upstream extremity of the patches is located at $s\approx 0.1$. Quantitative discrepancies between experimental and simulation results are, as mentioned previously, due to errors in the estimation of the plate rigidity $B$ used in our simulation.  Although an effort has been made during the experiments in order to reduce the impact of the electrode sectioning on the plate rigidity, this effect exists, and it is not accounted for in the theoretical model of the plate. Discrepancies in the absolute efficiency values between the two experiments have their origin  in the difference between coupling coefficients of each experiment: $\alpha=0.073$ for experiment 1 and $\alpha=0.062$ for experiment 2.
 
The good agreement between experiments and simulations also validates the model presented in section \ref{sec:model} and motivates its use in  the parametric study of more complex geometries presented in the remainder of this section. 

% ---------------------------
\subsection{Parametric study}

\begin{figure}[t]
\centering
\includegraphics[width=0.90\textwidth]{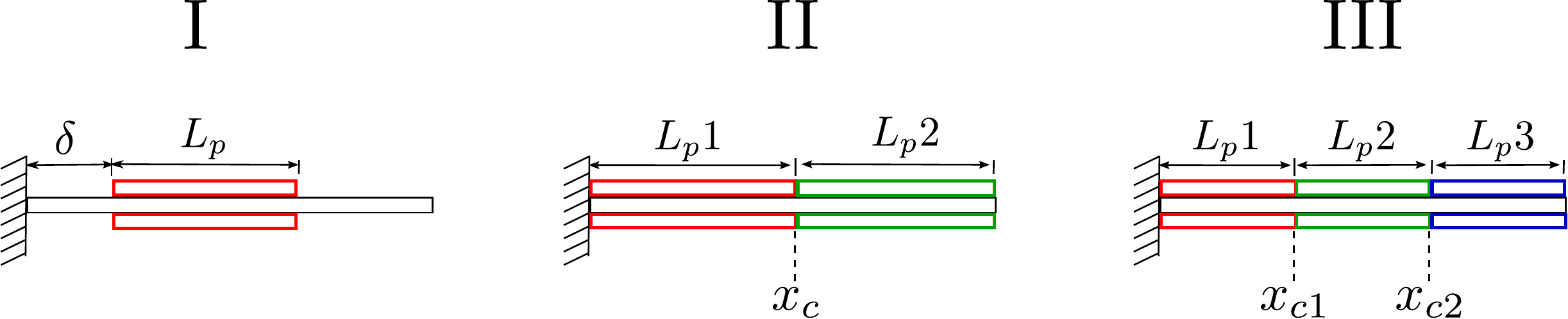}
\caption{Three cases of study in the energy harvesting efficiency of the piezoelectric flag.}
\label{piezo-dist}
\end{figure}

\noindent In this subsection, simulation results for the energy harvesting efficiency are presented in the three cases sketched in figure \ref{piezo-dist}:
\begin {description}
\item {I}: One single electrode with variable length $L_p$ and variable position along the flag surface (two-parameter problem)
\item {II}: Two electrodes of variable length covering the whole surface of the flag (one-parameter problem)
\item {III}: Three electrodes of variable length covering the whole surface of the flag (two-parameter problem)
\end{description}
For each case, we analyze the influence of size and distribution of the electrodes on the energy harvesting efficiency, defined as in the experimental case and expressed in non-dimensional form in Equation \eqref{efficacite}. 

A strong dependence of the efficiency on the system's physical parameters such as the mass ratio $M^*$, $\beta$, $U^*$ and  $\alpha$ is observed \cite{michelin2013}. Before exhibiting results of geometry optimizations, it is necessary to select a representative set of these parameters. First of all, due to the weakly-nonlinear nature of the model used in the present work, the accuracy of the results is restricted to flow velocities $U^*$ near the flapping instability threshold. Far from threshold this model exhibits discrepancies with previous simulations and a fully-nonlinear model of the system should be considered \cite{michelin2013}. Thus, all the results presented further in this work correspond to near threshold situations ($U^*$ is close to $U^*_c$), and hence, depend on the value of $M^*$ selected.

Secondly, two typical values of the mass ratio $M^*$ will be considered: $M^*=0.6$, which corresponds to a plate in a low density  fluid (\textit{i.e.} flag in air), and $M^*=10$ which corresponds to a plate in a dense fluid (\textit{i.e.} light plate in water). For each case, the flow velocity $U^*$ is adjusted to a typical value just above the instability threshold: $U^*=13$ and $8.25$, respectively. 

\begin{figure}
\centering
\includegraphics[width=0.60\textwidth]{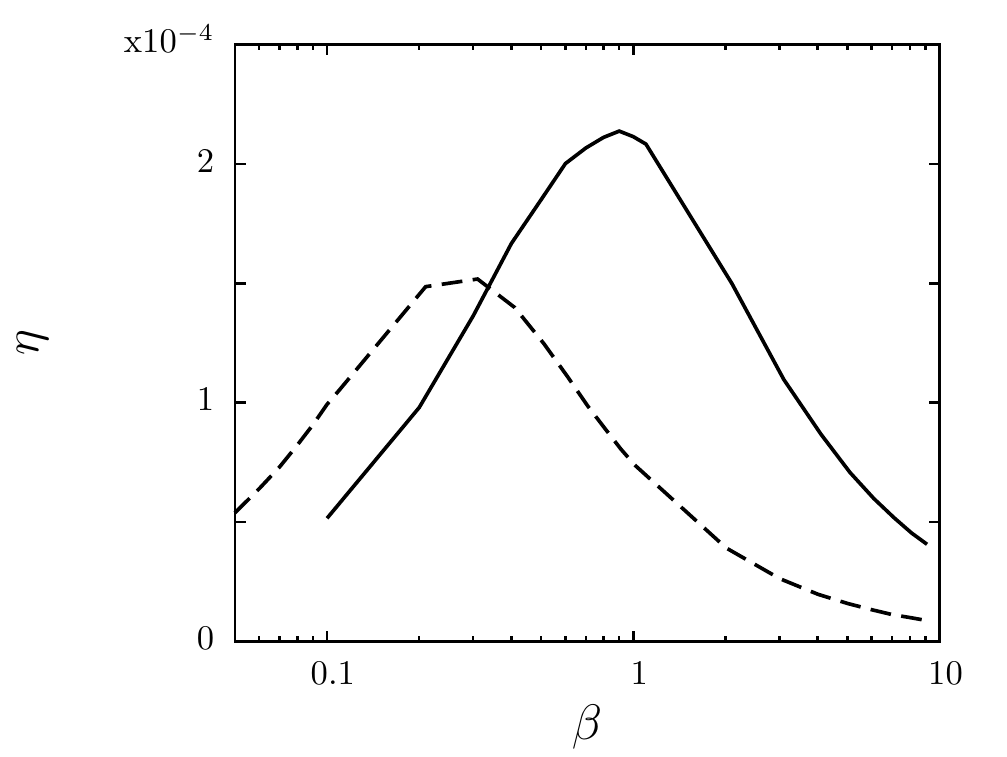}
\caption{Energy harvesting efficiency of a single electrode with $L_p=1$ as a function of $\beta$ for $M^*=0.6$, $U^*=13$ and $\alpha=0.25$ (solid line) and $M^*=10$, $U^*=8.25$ and $\alpha=0.25$ (dashed line).}
\label{effvsbeta}
\end{figure}

As explained in the previous section, maximum energy harvesting is achieved when the flapping timescale and the RC circuit timescale compare. Hence, the optimal value of $\beta$ will depend on the flapping frequency, which in turn depends on the mass ratio $M^*$ \cite{guo2000}. On figure \ref{effvsbeta} the efficiency $\eta$ in a typical single electrode configuration is plotted as function of $\beta$ for two different values of $M^*$. One then observes that higher values of $M^*$ tend to reduce the optimal value of $\beta$. Indeed, while for $M=0.6$ the optimal tuning parameter is $\beta\sim1$ (as in the experimental case), for $M^*=10$ the value of $\beta$ is about $0.3$. In the following, these optimal values for $\beta$ are retained. 

The influence of the coupling coefficient $\alpha$, the electrode position and geometry on the flapping frequency has then been tested. No significant impact on the flapping frequency, and consequently on the optimal value of $\beta$ has been observed. This is expected to result from the relatively low  coupling coefficients $\alpha$ considered.

\subsubsection*{Case I : single electrode}

\begin{figure}
\centering
\begin{tabular}{c@{}c}
\subfigure[$\eta$\quad($\alpha=0.1$)]{\includegraphics[width=0.5\textwidth]{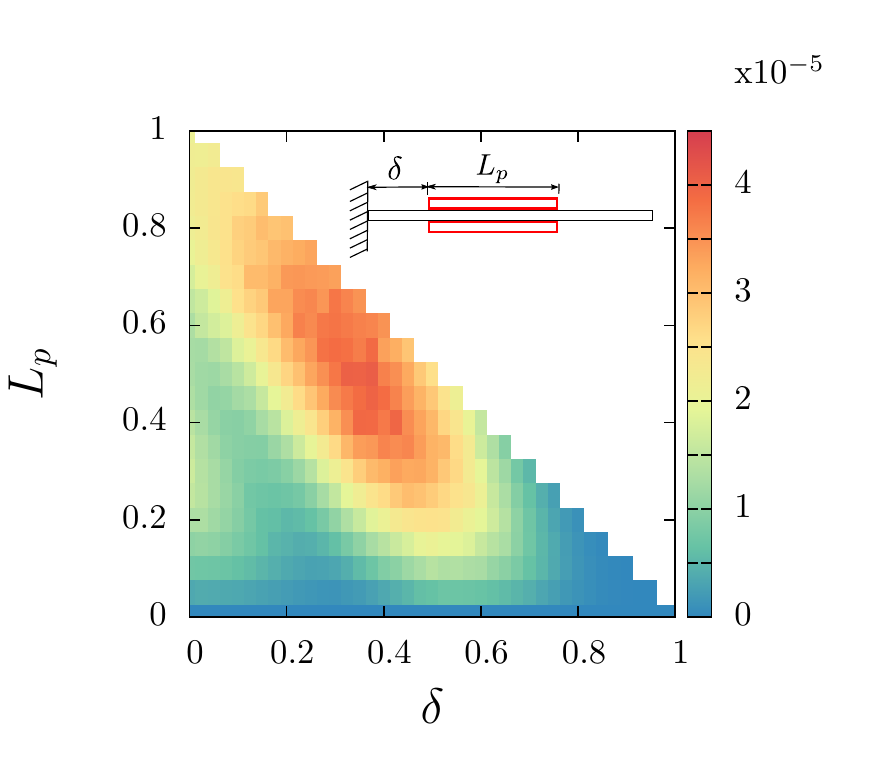}} &
\subfigure[$\eta$\quad($\alpha=0.3$)]{\includegraphics[width=0.5\textwidth]{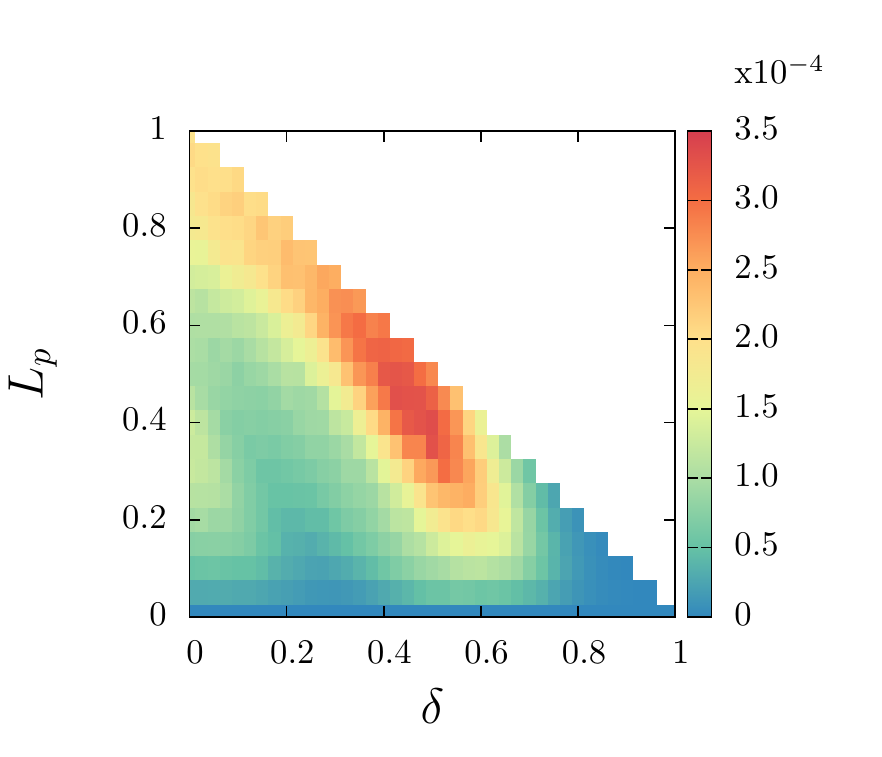}} \\
\multicolumn{2}{c}{\subfigure[$\eta$\quad($\alpha=0.5$)]{\includegraphics[width=0.5\textwidth]{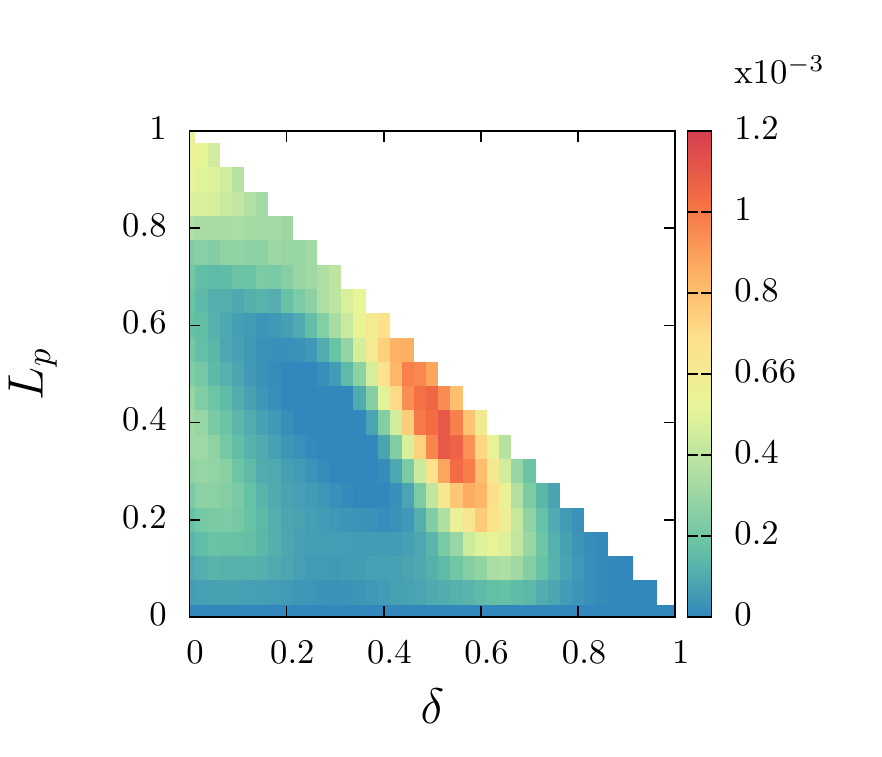}}}
\end{tabular}
\caption{Colormaps of the efficiency $\eta$ in the $(L_p,\delta)$ plane; for all cases, $M^*=0.67$, $U^*$=13 and $\beta=1$.\label{fig:effcase1_1}}
\end{figure}

\noindent We consider here the case  of one electrode  of length $L_p$ that  partially covers the surface of the plate. It is attached at a distance $\delta$ from the clamped end. Figure \ref{fig:effcase1_1} presents the efficiency of the harvesting system as a function of $\delta$ and $L_p$, for fixed values of $M^*=0.67$, $U^*=13$ and $\beta=1$, and increasing values of the electro-mechanical coupling factor $\alpha$. Each colored rectangle of these plots corresponds to one single numerical simulation.

We observe that regardless of the value of $\alpha$, the maximum efficiency is found for values of $\delta\neq 0$  and electrode lengths $0.35<L_p<0.5$.  In these cases, the electrode is mainly located  on the downstream half of the clamped-free plate. Moreover, the maximum efficiency region is observed to change as the electro-mechanical coupling factor $\alpha$ is increased. For small values of $\alpha$ ($\sim0.1$), the maximum efficiency region on the $(L_p,\delta)$ space extends from long electrodes $L_p=0.8$ placed at $\delta=0.2$ to smaller electrodes of length $L_p\sim0.4$ and $\delta\sim0.45$ (Figure \ref{fig:effcase1_1}a). As the coupling factor $\alpha$ is increased the maximum efficiency is also increased.  Simultaneously,  the maximum efficiency region shrinks and migrates in the $(L_p,\delta)$ plane to a new optimal configuration for an electrode of length $L_p\sim0.4$ and $\delta\sim0.55$ (Figure \ref{fig:effcase1_1}c). 

\begin{figure}[]
\centering
\begin{tabular}{cc}
%(a) & (b) \\
\subfigure[$\Delta A_N\quad(\alpha=0.1)$]{\includegraphics[width=0.5\textwidth]{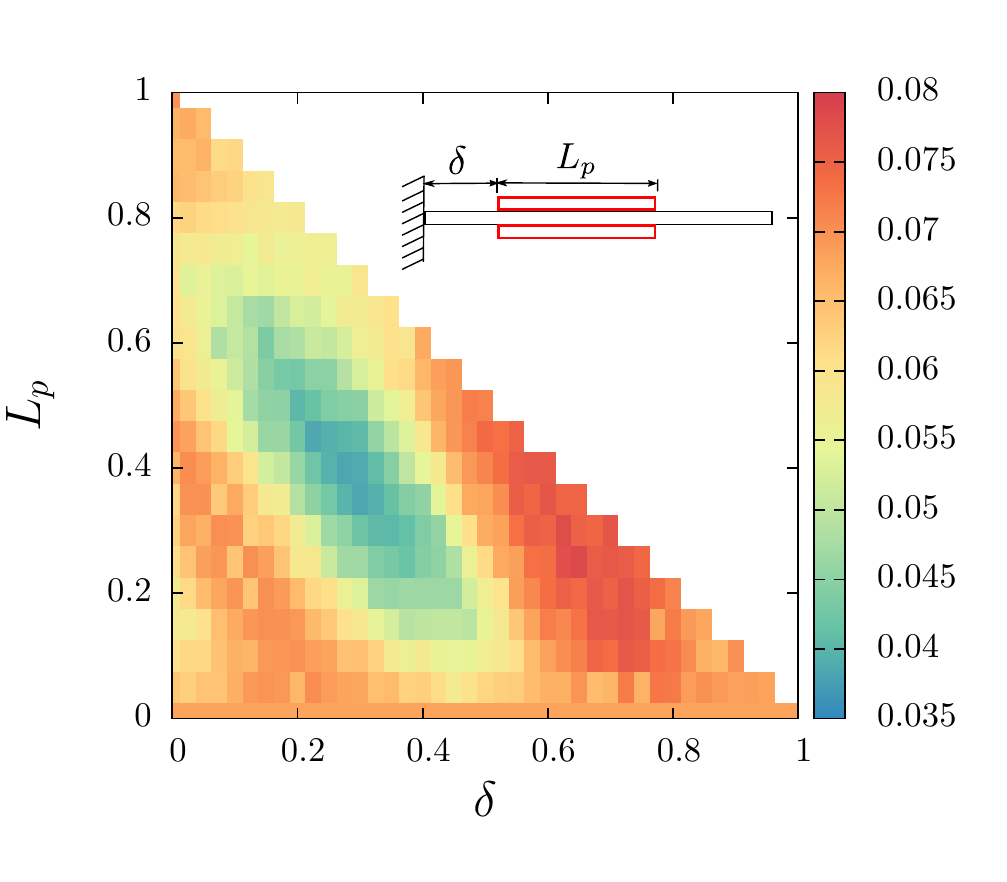}} &
\subfigure[$\Delta A_N\quad (\alpha=0.5)$]{\includegraphics[width=0.5\textwidth]{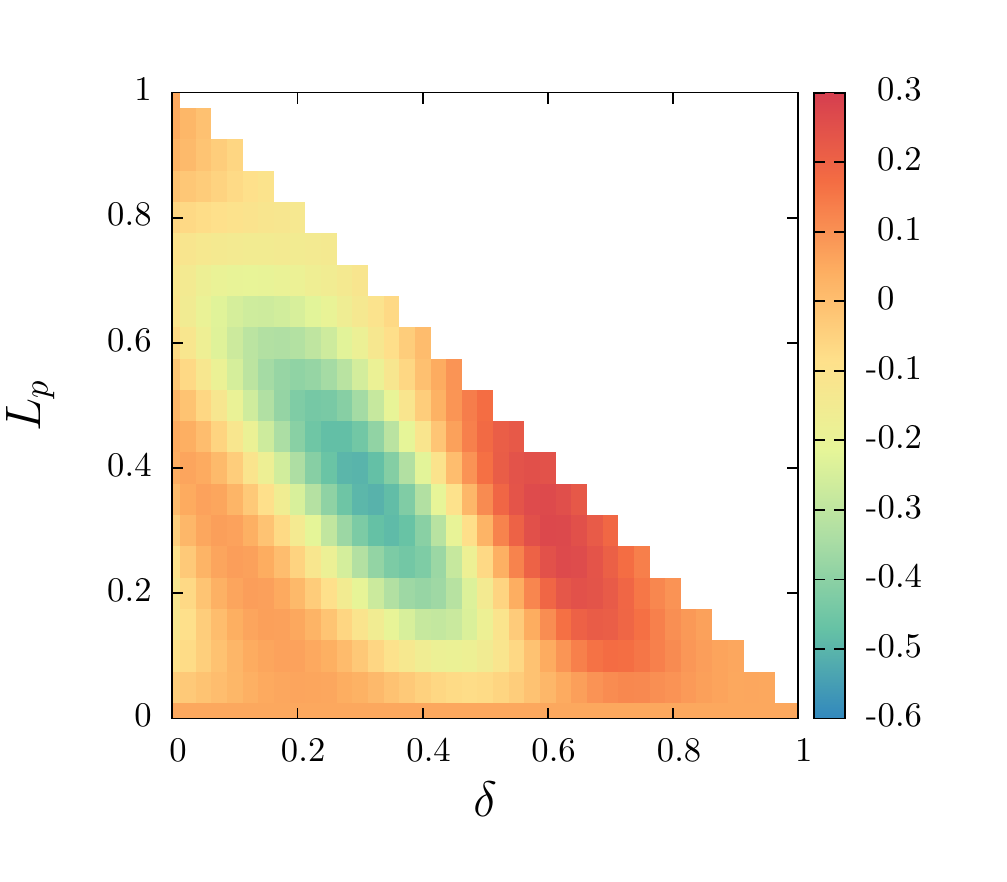}} \\
%(c) & (d) \\
\subfigure[$\Delta U_N^* \quad(\alpha=0.5)$]{\includegraphics[width=0.5\textwidth]{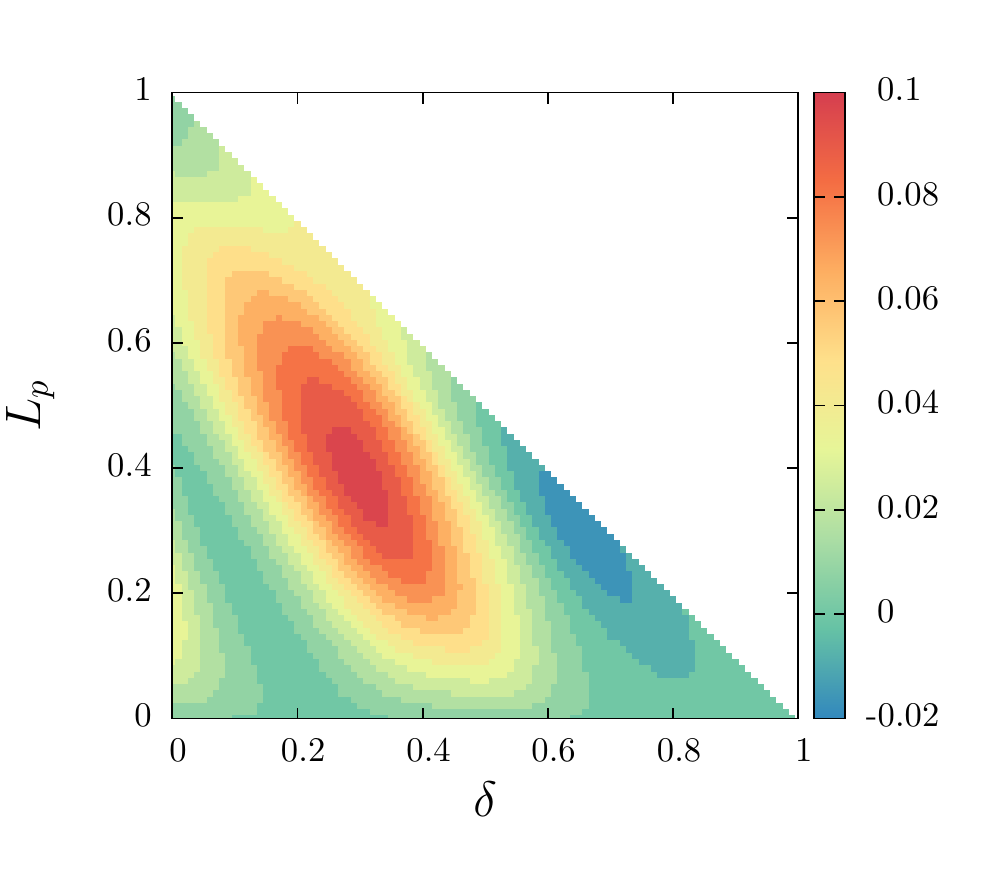}} &
\subfigure[$\overline{\Delta\theta_i} \quad (\alpha=0)$]{\includegraphics[width=0.5\textwidth]{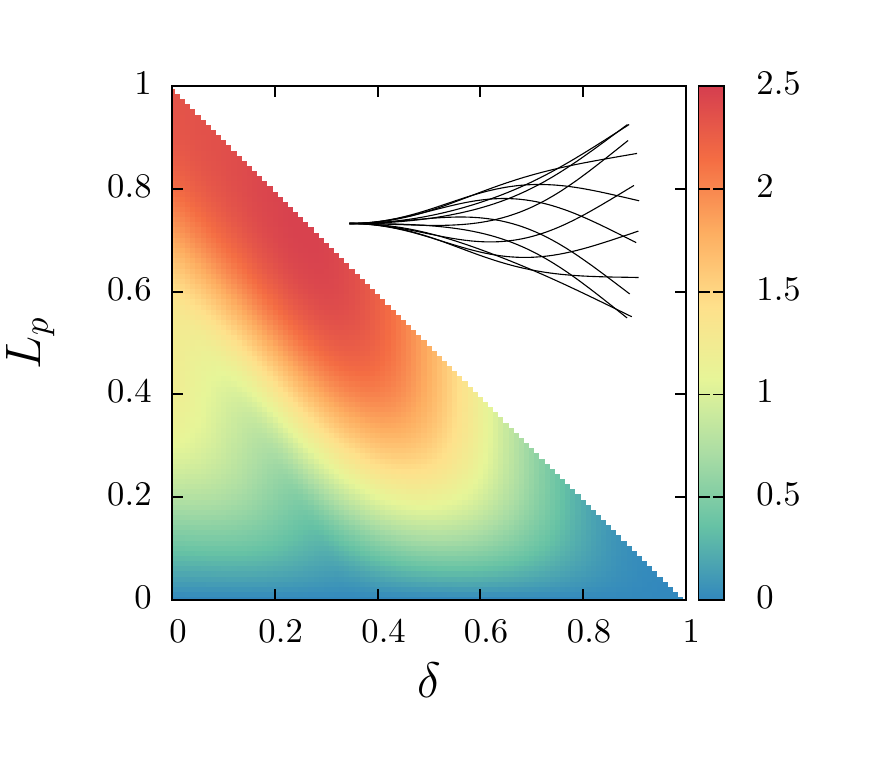}}
\end{tabular}
\caption{Normalized flapping amplitude difference $\Delta A_N=(A-A_0)/A_0$ in the $(L_p,\delta)$ plane for (a) $\alpha=0.1$ and (b) $\alpha=0.5$ respectively, where $A_0$ corresponds to the flapping amplitude with no coupling ($\alpha=0$).  (c) Normalized critical fluid velocity $\Delta U_N^*=(U_c^*-U_{c0}^*)/U_{c0}^*$ as a function of the electrode position in the $(L_p,\delta)$ plane where $U_{c0}$ corresponds to the flapping critical flow velocity with no coupling ($\alpha=0$). (d) $\overline{\Delta\theta_i}$ integrated over one oscillation cycle for a plate with no piezo-electrical coupling. The mode shape of the fluttering flag over one oscillation cycle is depicted in the inset. Other parameters are $M^*=0.67$, $U^*$=13 and $\beta=1$.}
 \label{fig:effcase1_2}
\end{figure}

To explain this difference in the optimal configuration, the influence of the geometrical parameters $(\delta,L_p)$ on the flapping amplitude is represented on figure \ref{fig:effcase1_2}(a,b) for small and large values of $\alpha$. These figures show the normalized relative flapping amplitude $\Delta A_N=(A-A_0)/A_0$ in the $(\delta,L_p)$ plane for $\alpha=0.1$ and $\alpha=0.5$, respectively, where $A_0$ is the flapping amplitude obtained with no coupling ($\alpha=0$). Comparison of these figures show that $\alpha$ only impacts the general magnitude of the amplitude fluctuation, and does not modify the parameter values leading to maximum (or minimum) amplitude. In particular, the maximum flapping amplitude corresponds to the optimal efficiency region achieved for high coupling. At large coupling, maximizing the flapping amplitude seems therefore to be the optimal strategy. These maximum flapping amplitudes are also associated with the largest reduction of the critical flow velocity (Figure \ref{fig:effcase1_2}c) which is consistent with the generally observed result that the flag flapping amplitude is a monotonically increasing function of $U^*-U^*_c$ in the vicinity of the threshold. 

On the other hand, when the coupling coefficient is small, the flapping flag dynamics is only marginally modified by the energy transfers to the piezoelectric patches. In that case, and neglecting the feedback coupling of the piezoelectric, the amplitude of the equivalent current generator (Figure \ref{esquema}a) can be directly obtained using Equation \eqref{eq_piezo_adim} from the kinematics of the flag \emph{in the case of no coupling}. More specifically it is proportional to the relative rotation rate $[\dot\theta]_{s_i^-}^{s_i^+}$ of the two edges of the piezoelectric patch. Maximum efficiency is therefore achieved for low coupling when $\overline{\Delta\theta_i}$ is maximum, where
\begin{equation}
\overline{\Delta\theta_i}=\left<|\dot\theta(s_i^+)-\dot\theta(s_i^-)|\right>,
\end{equation}
a result consistent with Figures \ref{fig:effcase1_1}(a) and \ref{fig:effcase1_2}(d).

Before presenting some results at a higher value of the mass ratio, let us summarize the results obtained so far:
\begin{itemize}\renewcommand{\labelitemi}{$-$}
\item{For weakly coupled systems ($\alpha=0.1$), the piezoelectric coupling has no influence on the flapping amplitude and the efficiency scales with $\overline{\Delta\theta_i}$ (figure \ref{fig:effcase1_2}d).}
\item{For strongly coupled systems $(\alpha=0.5)$, the piezoelectric coupling $\alpha$ has an influence on the flapping amplitude which in turn also influences the efficiency. It was finally shown that the effect of $\alpha$ on $\Delta A_N$ can be directly deduced from that of $\alpha$ on $\Delta U_N$.}
\end{itemize}

\begin{figure}
\centering
\begin{tabular}{cc}
\subfigure[Efficiency $\eta$ ($\alpha=0.25)$]{\includegraphics[width=0.5\textwidth]{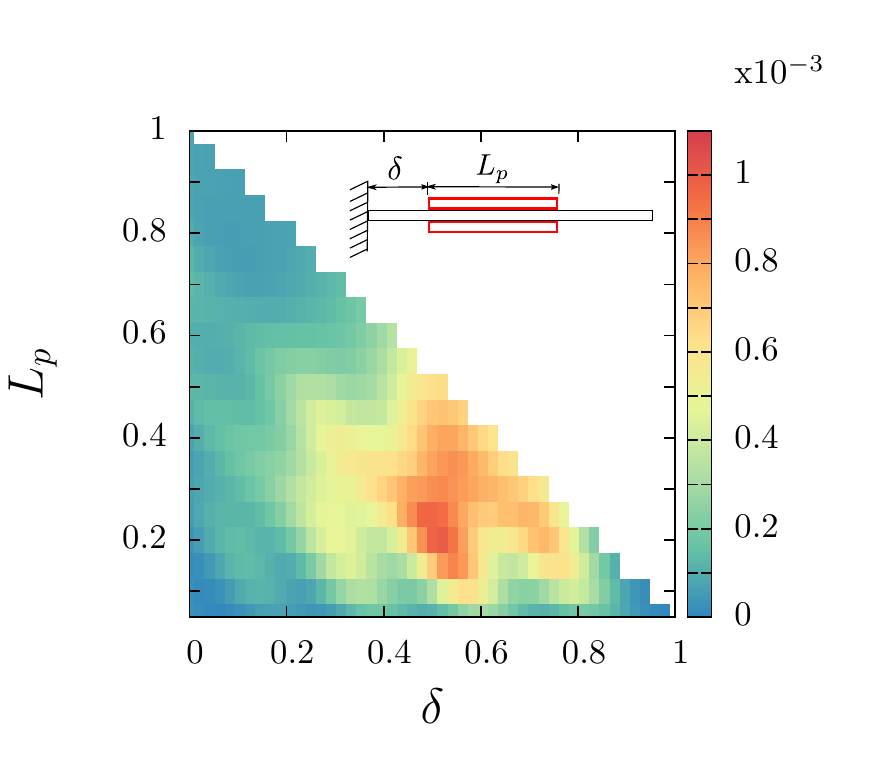}}&
\subfigure[$\Delta U_N^*\quad (\alpha=0.25)$]{\includegraphics[width=0.5\textwidth]{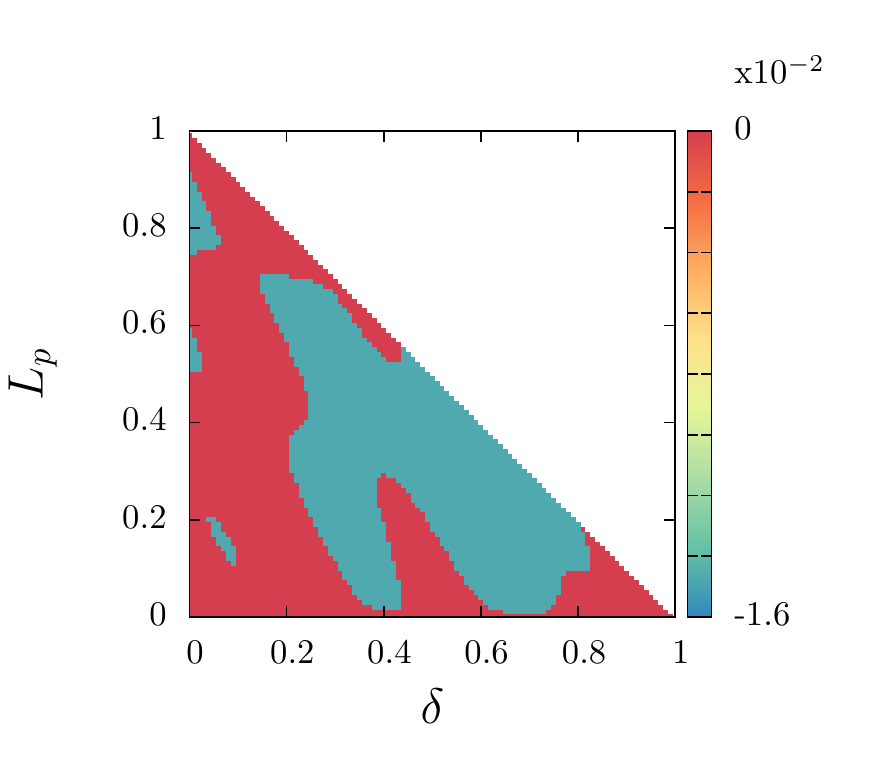}}\\
\multicolumn{2}{c}{\subfigure[$\overline{\Delta\theta_i}\quad(\alpha=0)$]{\includegraphics[width=0.5\textwidth]{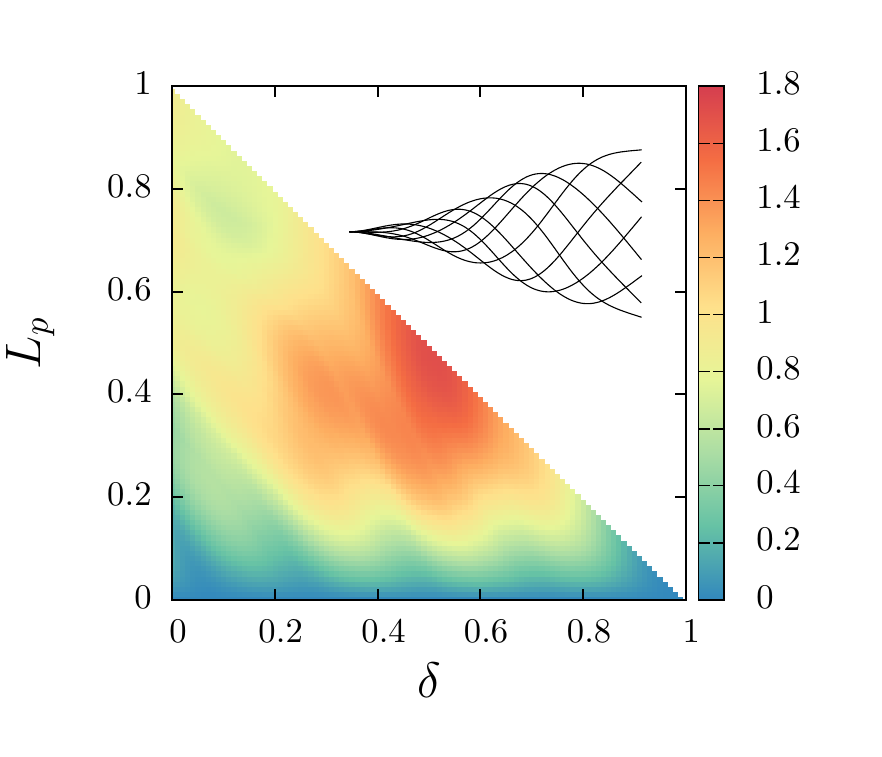}}}
\end{tabular}
\caption{ (a) Evolution of the efficiency $\eta$ with $(L_p,\delta)$ for $\alpha=0.25$, $M^*=10$, $U^*=8.25$ and $H^*=0.5$. (b) Normalized critical fluid velocity $\Delta U_N^*=(U_c^*-U_{c0}^*)/U_c^*$. (c) $\overline{\Delta\theta_i}$ integrated over one oscillation cycle for a plate with no piezo-electrical coupling. The mode shape of the fluttering flag over one oscillation cycle is depicted in the inset.}
\label{fig:effcase1_3}
\end{figure}

For $M^*=10$ and $\alpha=0.25$, we observe different local maxima of the efficiency in the ($\delta$,$L_p$) plane. They correspond to an electrode of length $L_p\sim0.2$ at $\delta \sim 0.3$, $0.5$ (absolute maximum) and $0.75$ respectively. The presence of several local maxima with shorter electrodes is explained by the fact that a larger $M^*$  leads to a higher order for the mode selected by the instability \cite{eloy2008}: each local maximum corresponds to the localization of the piezoelectric patches in the zones of maximum curvature. This is confirmed by the strong correlation of the maximum efficiency with $\overline{\Delta\theta_i}$ during the oscillating cycle (Figure \ref{fig:effcase1_3}c).

It should also be pointed out that the maximum efficiency for $M^*=10$ is about 2.5 times bigger than that for $M^*=0.67$. This result stems from the difference in mode structures at higher mass ratio: a shorter wave length penalizes long electrodes that only react to the average flag curvature along their length. It also suggests that for higher $M^*$  a larger number of smaller electrodes occupying the remaining free space on the plate surface could eventually increase the total efficiency even more.  This aspect is investigated further in the following. 

\subsubsection*{Case II: two  electrodes}
% ---------------------------------------

\begin{figure}
\centering
 \begin{tabular}{c @{}c}
\subfigure[$M^*=0.67$]{\includegraphics[width=0.5\textwidth]{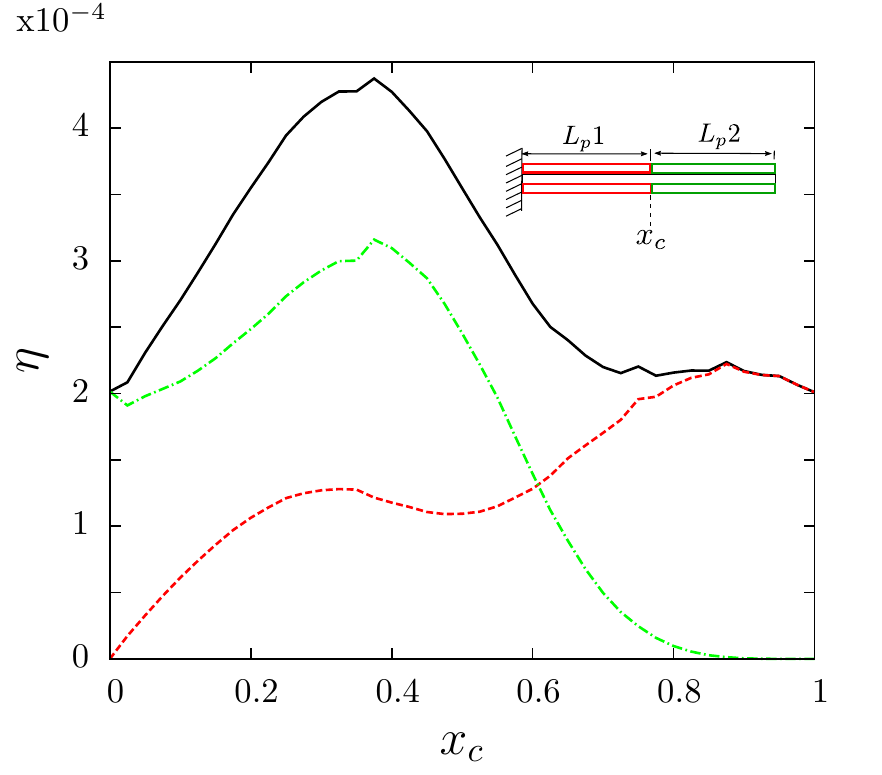}}&
\subfigure[$M^*=10$]{\includegraphics[width=0.5\textwidth]{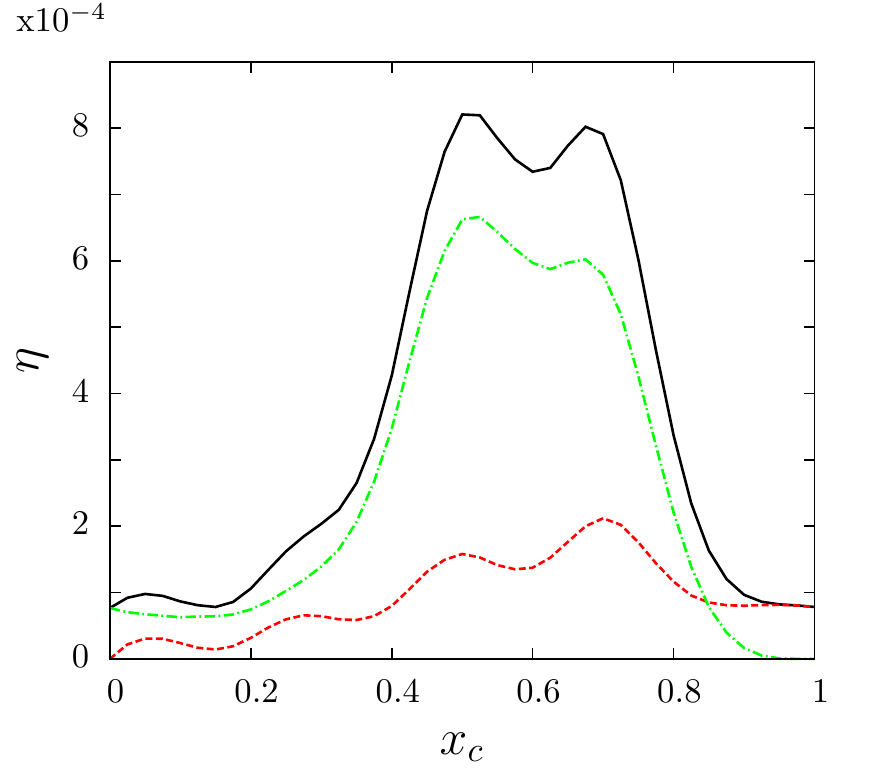}}\\
 \end{tabular}
 \caption{Energy harvesting efficiency for a piezoelectric plate covered by two electrodes of variables size for (a) $M^*=0.67$ and $U^*=13$, and (b) $M^*=10$ and $U^*=8.25$.  In both cases $\alpha=0.25$ and $H^*=0.5$. Each figure shows the total efficiency of the two electrode configuration (solid), as well as the efficiency of electrode 1 (dashed) and electrode 2 (dash-dotted).}
  \label{dos-piezo}
\end{figure}

\noindent The case of two electrodes covering the entire plate is now considered for the same values of $M^*$, $U^*$ and $\beta$ as in previous case, and depends on a single parameter namely the length of the first piezoelectric electrode $x_c$ (see figure \ref{piezo-dist}).  For $M^*=0.67$, the maximum efficiency is obtained for $x_c \sim 0.35$ (Figure \ref{dos-piezo} a) and is approximately twice as large as  the efficiency obtained for a single electrode covering the entire flag ($x_c=0$ or $x_c=1$), and represents a $6\%$ increase of the efficiency in comparison with the optimized one-electrode configuration detailed in Case I.  In this new optimal situation, the second electrode covers the most significant part of the plate's surface. Also, it is observed that for the same electrode's length, the downstream electrode leads to the largest energy harvesting, a result that is somehow expected because this region corresponds to the largest deflections of the flag.

For $M^*=10$, two local maxima are observed around at $x_c \sim 0.5$ and $x_c \sim 0.7$. Figure \ref{dos-piezo}(b) confirms that the downstream piezoelectric pair contributes the most to the efficiency for each maximum. However, the maximum efficiency of the two electrodes configuration is slightly smaller than that of case I, a consequence of the full coverage of the plate by the two electrodes, in particular near the leading edge where the local rotation vanishes. Due to the dominant mode structure at large $M^*$, shorter electrodes (either when using three or more electrodes or by choosing to cover only a fraction of the plate's surface) are therefore expected to improve again the efficiency.
 
\subsubsection*{Case III: three  electrodes}
% ------------------------------------------

\noindent We finally turn to the case of full coverage by three electrodes, described by the positions of the left edge of the first and second piezoelectric patches, $x_{c1}$ and $x_{c2}$ respectively.

\begin{figure}
\centering
 \begin{tabular}{l@{}l}
 (a)&(b)\\
\subfigure[$\eta$\quad($M^*=0.67$)]{\includegraphics[width=0.5\textwidth]{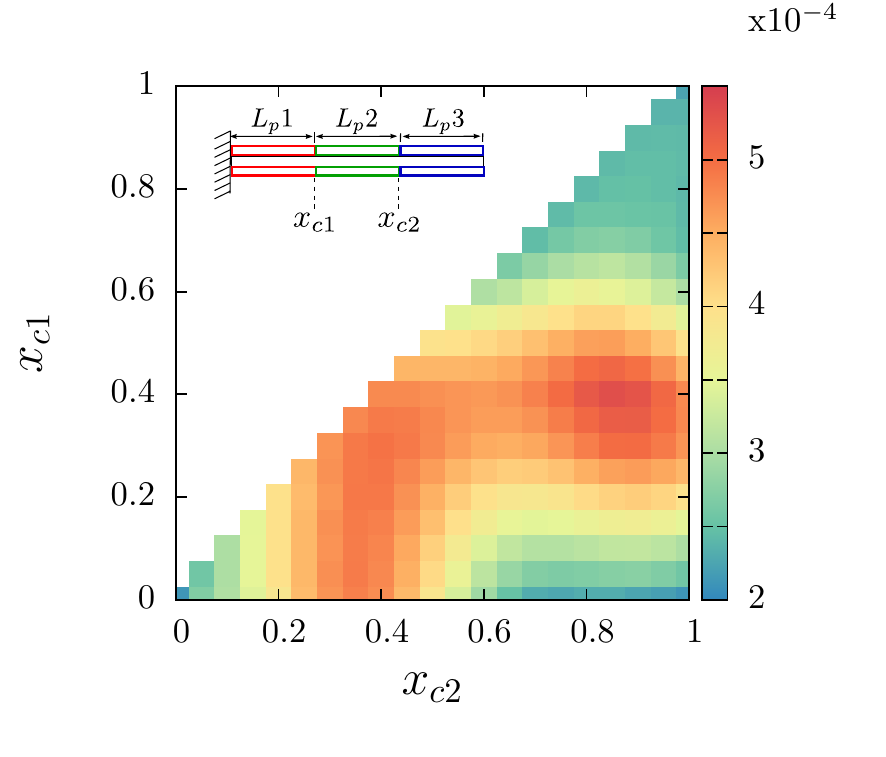}}&
\subfigure[$\eta$\quad($M^*=10$)]{\includegraphics[width=0.5\textwidth]{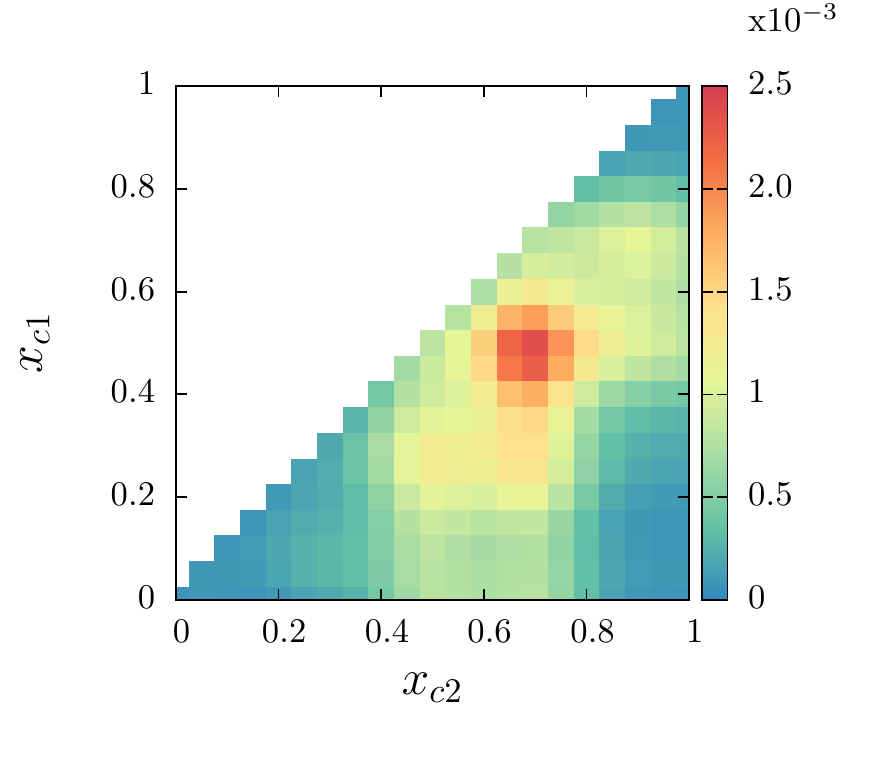}}
 \end{tabular}
 \caption{Energy harvesting efficiency for a piezoelectric plate covered by three electrodes of variables size.  (a) : $M^*=0.67$ and $U^*=13$.  (b) : $M^*=10$ and $U^*=8.25$.  In both cases $\alpha=0.25$ and $H=0.5$.}
\label{tres-piezos}
\end{figure}

For $M^*=0.67$ the maximum efficiency is reached for $x_{c1} \sim 0.4$ and $x_{c2} \sim 0.9$ (left plot in figure \ref{tres-piezos}). In this configuration, the second electrode is approximately the same as in the maximum efficiency configuration of case I. The presence of two additional electrodes results however in a $20\%$ increase of the efficiency in comparison with case I.  

For $M^*=10$, the arrangement of maximum efficiency $(x_{c1}=0.5,x_{c2}=0.7)$ is again related to the optimal configuration of case I, with the second electrode of length $L_p=0.2$ positioned at $\delta=0.5$. Taking advantage of the deformation mode, the third electrode of length $L_p=0.3$ and positioned at $\delta=0.7$ contributes in almost the same amount to the efficiency as the second electrode.  As a result, the total efficiency of the three electrode configuration is about three times higher than that of cases I and II.  

This clearly demonstrates that for higher mass ratios, a larger number of electrodes results in significant improvements of the efficiency, while this strategy only leads to moderate improvements for lower $M^*$.

%%%%%%%%%%%%%%%%%%%%%%%%%%%%%%%%%%%%%%%%%%%%%%%%%%%%%
\section {Conclusions}

\noindent This article investigated the possibility of converting kinetic energy of a flow into electrical energy through elastic deformation of a fluttering plate using piezoelectric effect, a problem that has received growing attention in recent years. A critical question lies in the optimization of the energy transfers between the fluid, solid and electric circuit. The present work focused on the role and influence of the location of the piezoelectric electrodes on the efficiency of the harvester, explicitly describing the discrete nature of the piezoelectric coverage of the flag and hence, allowing for a better representation of the experimental and practical situations.

To address this problem, the present work used a combination of theoretical analysis and experiments. A weakly nonlinear model considering a plate with a discrete number of piezoelectric patches pairs was developed and confronted to the experimental results. It was then used to determine geometries that maximize the energy harvesting efficiency in three different test cases: (1) one electrode partially covering the plate, (2) two electrodes fully covering the plate and (3) three electrodes fully covering the plate. For each configuration, the optimal geometries were described and analyzed.

The results presented in this work emphasize the critical role of the piezoelectric elements positioning: because the electric charge transfer responsible for the powering of the output circuit is solely given by the relative change of orientation of the flag between both ends of the patch, the placement of this patch must be carefully designed to take full advantage of the non-uniform flag deformation and its specificities, in particular the dependence of the mode shape with the mass ratio. In the case of a single electrode and small values of the mass ratio, the present study in fact shows that a greater amount of energy can be harvested when positioning the electrodes on the downstream half of the flag. At a higher value of the mass ratio, more than one optimal positioning are found, which is a consequence of more spatial oscillations of the unstable mode shape. Using a larger number of electrodes allows to improve the efficiency of the system particularly for higher mass ratios, where the modal structure corresponds to a shorter wavelength and is therefore better suited to shorter electrodes that are able to capture the flag's deformation best.

The work in this paper extends the results obtained in the context of damping or energy harvesting of vibrating plates in still fluid \cite{thomas2009,liao2012} to flow-induced instabilities and more than one piezoelectric patch. The present results allow to conclude that the optimal location for piezoelectric energy harvesting along the plate is on the downstream half, and in that regard are reminiscent of the conclusions of Ref.~\citep{singh2012b} obtained for a pure damping model of the energy harvesting process. In both cases, the local deformation of the flag is driving the energy harvesting process, either through changes in the local curvature or in the relative orientation at both ends of the piezoelectric patches, and the response of the deformation of the flag to the presence of energy harvesting plays a critical role in determining the optimal location for large coupling coefficient. Both approaches however differ in the relevant measure of flag deformation (either curvature or relative orientation), which also impacts the detail of the optimal location for the damping or piezoelectric patches.

In the case of weakly coupled systems, it was also shown that the amount of harvested energy can be directly predicted from the kinematics of the flapping flag obtained for zero piezoelectric coupling. Maximizing the harvested energy hence requires to determine the positioning that will ensure a maximum relative variation of the orientation between the two ends of the piezoelectric components. In the case of strongly coupled system, the positioning of the electrodes may influence the flapping amplitude, and consequently the efficiency, because of the feedback effect of the piezoelectric patches on the flag dynamics. Near the stability threshold, the flapping amplitude can be directly related to the critical velocity fluctuations: in that regard, linear stability analysis may be a great tool to estimate the impact of the piezoelectric coverage on the efficiency in the case of large coupling.

%%%%%%%%%%%%%%%%%%%%%%%%%%%%%%%%%%%%%%%%%%%%%%%%%%%%%
\section*{Ackowledgments}
This work was supported by the ``Laboratoire d'Excellence'' LASIPS (project PIEZOFLAG), and the French National Research Agency ANR (Grant ANR-2012-JS09-0017). S.M. also acknowledges the support of a Marie Curie International Reintegration Grant within the 7th European Framework Program (Grant PIRG08-GA-2010-276762).

%%%%%%%%%%%%%%%%%%%%%%%%%%%%%%%%%%%%%%%%%%%%%%%%%%%%%
\appendix
\section{}\label{Galerkin-projection}

The vertical displacement $y$ is expanded on clamped-free beam eigenmodes $\phi_p(s)$,

\begin{equation}
y=\sum_{i=1}^N\phi_i(s)q_i(t).
\end{equation}
After truncation to N linear modes, equation (\ref{ydest}) is projected along mode $p$. Projection of the forces $L(y)$, $f_\text{m}$, $f_\text{B}$, $f_\chi$, $f_{\text{res}}$ and $f_{\text{reac}}$ are given in this appendix. Projection of $L(y)$ along mode $p$ reads,
\begin{align}
L^p(y)&=\ddot{q}_p+\lambda_p^4q_p-\frac{\alpha}{U^*}\sum_i^{Np} v_i(\phi_p'(s_i^+)-\phi_p'(s_i^-))\nonumber\\
&+M^* H^*\left(\ddot{q}_p +\sum_{i=1}^Na_i^p q_p+2\sum_{i=1}^Nb_i^p\dot{q}_p\right),
\end{align}
where
\begin{eqnarray}
a_i^p=\int_0^1\phi_p\phi''_i ds,\\
b_i^p=\int_0^1\phi_p\phi'_i ds.
\end{eqnarray}
 For $f_\text{m}(y)$ we have:
 \begin{equation}
 f^p_\text{m}(y)=\sum_{i,j,k=1}^N(q_i\dot{q}_j\dot{q}_k+q_i q_j\ddot{q}_k)(c_{i,j,k}^p+d_{i,j,k}^p),
 \end{equation}
 with
 \begin{align}
 c_{i,j,k}^p&=\int_0^1\phi_p\phi'_i\int_0^{s'}\phi'_j\phi'_k ds ds',\\
 d_{i,j,k}^p&=\int_0^1\phi_p\phi''_i\int_{1}^{s'}\int_0^{s''}\phi'_j\phi'_k ds ds''ds'.
 \end{align}
The term  $f_\text{B}(y)$ projects as:
\begin{equation}
f^p_\text{B}(y)=\sum_{i,j,k=1}^Nq_i q_j q_k(e_{i,j,k}^p+4f_{i,j,k}^p+g_{i,j,k}^p),
\end{equation}
where 
\begin{align}
e_{i,j,k}^p&=\int_0^1\phi_p\phi''_i\phi''_j\phi''_k ds,\\
f_{i,j,k}^p&=\int_0^1\phi_p\phi'_i\phi''_j\phi'''_k ds,\\
g_{i,j,k}^p&=\int_0^1\phi_p\phi'_i\phi'_j\phi^{(4)}_k ds.
\end{align}
The mechanical-electrical coupling term $f_\chi$ gives
\begin{align}
f^p_\chi(y,v_1..{v_N}_p)&=\frac{1}{2}\sum_{j,k=1}^N q_jq_k\sum_{i=1}^{N_p}v_i\left(A_{j,k}^p(s_i^+)-A_{j,k}^p(s_i^-)\right),
\end{align}
with
\begin{align}
A_{j,k}^p=\phi'_j\phi'_k\phi'_p+\phi''_j\phi'_k\phi_p+\phi'_j\phi''_k\phi_p-2\phi_p\phi'_j\phi''_k.
\end{align}
For the resistive term $f_{res}$ we have :
\begin{equation}
f^p_{\text{res}}(y)=\frac{1}{2}C_Dt_{j,k}^p,
\end{equation}
where
\begin{equation}
t_{j,k}^p=\int_0^s\phi_p\sum_{i,j=1}^N|\phi'_i q_i +\phi_i \dot{q}_i|(\phi'_j q_j +\phi_j \dot{q}_j)ds.
\end{equation}

Notice that unlike the other coefficients $t_{j,k}^p$ is a function of time. It is due to the presence of the absolute value. This coefficient must then be recalculated at each time step.

Finally, for $f_{\text{reac}}$ we have:
\begin{align}
f^p_{\text{reac}}(y)&=-\sum_{i,j,k=1}^N\left(\frac{1}{2}\beta_{i,j,k}^p+\varpi_{i,j,k}^p\right)q_i q_j q_k+\sum_{i,j,k=1}^N(\eta_{i,j,k}^p-3\zeta_{i,j,k}^p-2\varsigma_{i,j,k}^p) q_i q_j \dot{q}_k\nonumber\\
&-2\sum_{i,j,k=1}^N\xi_{i,j,k}^p \dot{q}_i \dot{q}_j q_k-\frac{1}{2}\sum_{i,j,k=1}^N\sigma_{i,j,k}^p q_i \dot{q}_j \dot{q}_k+\sum_{i,j,k=1}^N\psi_{i,j,k}^p (q_i \ddot{q}_j q_k+q_i \dot{q}_j \dot{q}_k+2 \dot{q}_i \dot{q}_j q_k)\nonumber\\
&+2\sum_{i,j,k=1}^N\varphi_{i,j,k}^p q_i \dot{q}_j q_k-\sum_{i,j,k=1}^N\varrho_{i,j,k}^p q_i q_j \ddot{q}_k
\end{align}

with:
\begin{align}
\beta_{i,j,k}^p&=\int_0^1\phi_p\phi'_i\phi'_j\phi''_k ds,\\
\eta_{i,j,k}^p&=\int_0^1\phi_p\phi'_i\phi'_j\phi'_k ds,\\
\zeta_{i,j,k}^p&=\int_0^1\phi_p\phi'_i\phi''_j\phi_k ds,\\
\xi_{i,j,k}^p&=\int_0^1\phi_p\phi_i\phi'_j\phi'_k ds,\\
\sigma_{i,j,k}^p&=\int_0^1\phi_p\phi''_i\phi_j\phi_k ds,\\
\psi_{i,j,k}^p&=\int_0^1\phi_p\phi'_i\int_0^{s'}\phi'_j\phi'_k dsds',\\
\varphi_{i,j,k}^p&=\int_0^1\phi_p\phi''_i\int_0^{s'}\phi'_j\phi'_k dsds',\\
\varrho_{i,j,k}^p&=\int_0^1\phi_p\phi''_i\int_{s'}^{1}\phi'_j\phi_k dsds',\\
\varsigma_{i,j,k}^p&=\int_0^1\phi_p\phi''_i\int_{s'}^{1}\phi'_j\phi'_k dsds',\\
\varpi_{i,j,k}^p&=\int_0^1\phi_p\phi''_i\int_{s'}^{1}\phi'_j\phi''_k dsds'.
\end{align}

%==================================================================
\bibliographystyle{unsrt}

\end{document}